\documentclass[sn-mathphys]{sn-jnl}
\usepackage[utf8]{inputenc}
\usepackage[T1]{fontenc}
\usepackage{graphicx}
\usepackage{grffile}
\usepackage{longtable}
\usepackage{rotating}
\usepackage{xcolor}
\usepackage[normalem]{ulem}
\usepackage{amsmath}
\usepackage{textcomp}
\usepackage{amssymb}
\usepackage{dsfont}
\usepackage{hyperref}
\usepackage{enumitem}
\setlist{noitemsep}

\newcommand{\yo}[1]{{\color{black} #1}}

\newcommand{\bp}[0]{\ensuremath{\boldsymbol{p}}}
\newcommand{\bx}[0]{\ensuremath{\boldsymbol{x}}}
\newcommand{\by}[0]{\ensuremath{\boldsymbol{y}}}
\newcommand{\bq}[0]{\ensuremath{\boldsymbol{q}}}
\newcommand{\bk}[0]{\ensuremath{\boldsymbol{k}}}

\begin{document}

\title[Large deviations for linear wave kinetic equation]{Dynamical large deviations for an inhomogeneous wave kinetic theory: linear wave scattering by a random medium}

\author*[1,2]{\fnm{Yohei} \sur{Onuki}}\email{onuki@riam.kyushu-u.ac.jp}
\author[2]{\fnm{Jules} \sur{Guioth}}
\author[2,3]{\fnm{Freddy} \sur{Bouchet}}

\affil[1]{\orgdiv{Research Institute for Applied Mechanics}, \orgname{Kyushu University}, \orgaddress{\city{Kasuga}, \state{Fukuoka}, \country{Japan}}}

\affil[2]{\orgname{Laboratoire de Physique, ENS de Lyon, CNRS}, \orgaddress{\city{Lyon}, \country{France}}}

\affil[3]{\orgname{LMD/IPSL, ENS, Université PSL, École Polytechnique, Institut Polytechnique de Paris, Sorbonne Université, CNRS}, \orgaddress{\city{Paris}, \country{France}}}

\abstract{The wave kinetic equation predicts the averaged temporal evolution of a continuous spectral density of waves either randomly interacting or scattered by the fine structure of a medium. In a wide range of systems, the wave kinetic equation is derived from a fundamental equation of wave motion, which is symmetric through time-reversal. By contrast, the corresponding wave kinetic equations is time-irreversible: its solutions monotonically increase an entropy-like quantity. A similar paradox appears whenever one make\yo{s} a mesoscopic description of the evolution of a very large number of microscopic degrees of freedom, the paradigmatic example being the kinetic theory of dilute gas molecules leading to the Boltzmann equation. Since Boltzmann, it has been understood that a probabilistic understanding solves the apparent paradox. More recently, it has been understood that the kinetic description itself, at a mesoscopic level, should not break time reversal symmetry \cite{bouchet2020boltzmann}. The time reversal symmetry remains a fundamental property of the mesoscopic stochastic process: without external forcing the path probabilities obey a detailed balance relation with respect to an equilibrium quasipotential. The proper theoretical or mathematical tool to derive fully this mesoscopic time reversal stochastic process is large deviation theory: a large deviation principle uncovers a time reversible field theory, characterized by a large deviation Hamiltonian, for which the deterministic wave kinetic equation appears as the most probable evolution. Its irreversibility appears as a consequence of an incomplete description, rather than as a consequence of the kinetic limit itself, or some related chaotic hypothesis. This paper follows \cite{bouchet2020boltzmann} and a series of other works that derive the large deviation Hamiltonians of the main classical kinetic theories, for instance \cite{GBE}  for homogeneous wave kinetics. We propose here a derivation of the large deviation principle in an inhomogeneous situation, for the linear scattering of waves by a weak random potential. This problem involves microscopic scales corresponding to the typical wavelengths and periods of the waves and mesoscopic ones which are the scales of spatial inhomogeneities in the spectral density of both the random scatterers and the wave spectrum, and the time needed for the random scatterers to alter the wave spectrum. The main assumption of the kinetic regime is a large separation of these microscopic and mesoscopic scales. For the sake of simplicity, we consider a generic model of wave scattering by weak disorder: the Schr\"odinger equation with a random potential. We derive the path large deviation principle for the local spectral density and discuss its main properties. We show that the mesoscopic process obeys a time-reversal symmetry at the level of large deviations.}

\maketitle



This publication is part of a special issue in homage of the memory of Krzysztof Gaw\c{e}dzki. The subject of this work is large deviation theory applied to wave turbulence. Large deviation theory applied to complex dynamics and turbulent flows was one of the subjects for which Krzysztof Gaw\c{e}dzki made a number of important contributions during the last few years, see for instance  \cite{cardy2008non,chetrite2007kraichnan,gawedzki2008stochastic,gawedzki2013fluctuation,chetrite2009eulerian,bouchet2016perturbative}. He taught many of us, including Freddy Bouchet, many aspects of large deviation theory. We wrote a common paper on the subject of large deviation theory and non-equilibrium quasipotentials for stochastic particles with mean field interactions \cite{bouchet2016perturbative}. Given his scientific qualities, and his deep sense of friendship, it is great pleasure for us to pay homage to Krzysztof Gaw\c{e}dzki through this modest contribution.

\section{Introduction}\label{sec1}

The aim of this paper is to extend the existing kinetic theory to describe probabilistically mesoscopic evolutions of wave fields interacting with random potentials. We will derive for the first time a large deviation principle that describes completely typical and rare fluctuations of the wave local spectral density. This is also the first large deviation principle for wave kinetic theory in an inhomogeneous setup. This work lies at the intersection of three different active fields in theoretical and mathematical physics: the description of waves interacting with random media and their applications to ocean and atmosphere dynamics, recent mathematical and theoretical advances in the kinetic theory of wave turbulence, and large deviation theory for kinetic theories.

For the first field, we note that wave propagation in disordered media is a ubiquitous phenomenon appearing in various areas of physics. Typical examples include light radiation through the atmosphere, acoustic or internal gravity waves in turbulent flows, and elastic waves in solid Earth. In most cases, one is not interested in the individual wave interference or scattering processes but in the statistical description of the overall wave field at a mesoscopic spatial scale much greater than the extent of disorder or wavelength. For this purpose, it is customary to define the spectral density of the wave signal at each location and investigate its statistical properties. The wave kinetic equation, sometimes referred to as a radiative transport equation or simply as a transport equation, is known as the universal model to describe the evolution of the local spectral density. It commonly derives from elementary wave equations and has a broad range of applications \cite{ryzhik1996transport,bal2005kinetics,powell2005transport,danioux2016near,savva2018scattering,savva2021inertia}. Recently, the evolution of wave spectra under scattering interactions with a turbulent flow were studied in a two-dimensional model \cite{BBS}.

Wave kinetic theory is of special interest in some specific areas of ocean and atmosphere research. Since the celebrated work by Klaus Hasselmann \cite{hasselmann1962non}, the kinetic description of nonlinear 4-wave interactions among water waves has been used for estimating energy transfer rates in a wind wave spectrum and forecasting the sea surface states. The linear counterpart of the wave kinetic equation is relevant to surface or internal wave energy dispersion in a slowly evolving turbulent flow \cite{danioux2016near,savva2018scattering,kafiabad2019diffusion,boas2020directional,dong2020frequency}. In these actual problems, the scale-separation assumption at the heart of the kinetic theory might be valid, but is not necessarily always valid. For example, internal wave activity in the ocean is highly heterogeneous, which is imprinted on the variability of energy dissipation rates on scales of order 10 to 100 km, in the mid-depth layer \cite{whalen2015estimating}. For tide or wind generated waves with $10$-$100$ km horizontal wavelengths, deviations of the spectral evolutions from that predicted by the kinetic equation may not be negligible, and a first principle theory of fluctuation is missing. This motivates us to revisit the theoretical basis of wave kinetic theory. Our work can be considered as the first building block for stochastic parameterization of the local spectral density from first principles, for the specific case of wave interacting with random potentials.

In relation with geophysical applications, several experiments with fundamental scopes in wave kinetic theory have been recently devised. For instance this led to the very first observation of the regime of inertial wave turbulence in a rotating flow \cite{MBGC}, the identification of regimes of weakly and strongly nonlinear internal wave turbulence in an experiment of stratified turbulence \cite{RSDRASVVM}, experiments on statistical properties of water waves in a large basin \cite{Mich22}, the validation of the inverse cascade phenomenon \cite{Fal20}, and extension of the range of scales for observing pure gravity wave turbulence in the laboratory \cite{Caz19} using reduced gravity experiments.

The second field, fundamental theoretical developments in wave turbulence theory, has seen many new advances recently. For the first time, using numerical simulations of the non-linear Schrödinger wave kinetic equation, predictions by the wave kinetic equation were tested for several kinetic times \cite{BBKKS,ZSKN}. Novel finite-size effects in wave turbulence were systematically studied in a one-dimensional model using a combination of theory and numerics \cite{DB}. \yo{Initiating from the asymptotic statistics of the non-linear Schr\"odinger equation in an equilibrium state \cite{lukkarinen2011weakly}, significant recent progress has been made to give a mathematical foundation for wave turbulence theory: theorem about approximations of the dynamics for times much shorter than the kinetic time \cite{BGHS,CG19,CG20,Faou,ACG}, the understanding of propagation of chaos \cite{DH22}, beginning investigation of water wave equation \cite{DIP}, and remarkable first full rigorous derivations of the wave kinetic theory at the kinetic timescale, for the non-linear Schr\"odinger equation \cite{DH19} and KdV type equations \cite{staffilani2021wave,hannani2022wave}.}
From the point of view of these fundamental perspectives, our work gives for the first time a description of all the cumulants of the local spectral density, through a large deviation principle, in an inhomogeneous setting.

The third field is the development of large deviation principles in relation with kinetic theory. Many classical equations of mathematical physics arise from a law of large numbers, when faster and smaller scale degrees of freedom are averaged out. This is the case for all classical kinetic theories. It is natural to extend all these theories to look for the statistics of fluctuations. Generically, one expects to derive a large deviation principle that describes a statistical field theory quantifying the probabilities of any fluctuations, either typical or extremely rare, in a way analogous to macroscopic fluctuation theory \cite{bertini2015macroscopic} for stochastic diffusive systems, or large deviation theory for stochastic dynamics with mean field interaction \cite{bouchet2016perturbative}. Deriving such large deviation principles from deterministic microscopic dynamics is a fundamental endeavor in theoretical and mathematical physics. Recently, the large deviation principles for a number of classical kinetic theories, starting from first principles, have been uncovered: for discrete models that mimic dilute gases and with Boltzmann like behavior \cite{leonard1995large,rezakhanlou1998large}, for dilute gases related to the Boltzmann equation \cite{bouchet2020boltzmann,bodineau2020fluctuation}, for the Kac model \cite{heydecker2021large,basile2022asymptotic}, for plasma at length scales much smaller than the Debye length related to the Landau equation \cite{feliachi2021dynamical}, for homogeneous systems with long range interactions related to the Balescu--Guernsey--Lenard equation \cite{feliachi2022dynamical}, for weakly interacting waves in a homogeneous setup \cite{GBE} related to the wave kinetic equation. The large deviation principles describe fluctuations but also uncover gradient structure for the deterministic kinetic equation, see \cite{mielke2014relation} and a simple explanation in \cite{bouchet2020boltzmann}. Several mathematical results, usually valid for a fraction of the kinetic time in the spirit of Lanford results for the deterministic equation, have been obtained for the large deviation principles, for instance for the Boltzmann equation \cite{bodineau2020fluctuation,bodineau2020statistical}, or for the Kac model with unexpected corrections to the expected large deviation principle \cite{heydecker2021large,basile2022asymptotic} associated to giant concentrations and solutions of the Boltzmann equation without energy conservation.

One aim of large deviation theory is to study rare events. In the context of wave dynamics, large deviation theory has been used to study rare events for the evolution of the empirical spectrum \cite{GBE} on the kinetic time scale, but also for studying the appearance of very large amplitude waves \cite{DGOVE}, for the non-linear Schr\"odinger dynamics for shorter time scales. Instantons structures have been predicted and compared with experimental data taken from a 300 m long wave \cite{DGOVE}. Large deviation principles for the wave amplitude due to short time phase mixing has also been studied \cite{GGKS}.

The result described in this paper is a new example of a large deviation principle for a kinetic theory, derived from microscopic dynamics. It is the first extension of large deviation theory for the local spectral density in an inhomogeneous setup. It opens the way for other inhomogeneous large deviation principle for wave turbulence, and for the study of new wave turbulence phenomena where rare events play an important role.

As a generic model of waves interacting with random medium, we consider the linear Schr\"odinger equation in a weak random potential
\begin{align*}
i \frac{\partial \psi}{\partial t} = - \frac{D}{2} \nabla_x^2 \psi + V \psi,
\end{align*}
where $\psi(\boldsymbol{x}, t)$ is a wave function defined on $\mathbb{R}^{d+1}$ and $V(\boldsymbol{x})$ is a homogeneous random potential. For this model, we assume a regime with a wave spectrum which is dominated by waves of typical wavelengths $\lambda$, and with modulations of the statistical properties of the wave spectrum on scales of order $\lambda / \mu$. The second assumption is that the typical correlation length of the potential is of order $\lambda$ and that interactions between the waves and the potential is weak, more precise definitions are given in section \ref{sec:WKE}. Then, for small value of $\mu$, we have a separation of scales and of the associated times, where a huge amount of waves experience multiple scattering in domains of typical size $\lambda / \mu$. It is natural to focus on variations of the field on the mesoscopic scales of order  $\lambda / \mu$. This defines a kinetic regime where such mesoscopic variations are captured by the Wigner distribution $n$, that somehow measures the wave energy density in both position and wave-vector space. After time and length rescaling $t \to \mu t$ and $\boldsymbol{x} \to \mu \boldsymbol{x}$, in the small $\mu$ limit, the wave kinetic equation is classically derived (see for instance \cite{ryzhik1996transport,erdHos2000linear}):
\begin{align*}
\frac{\partial n(\boldsymbol{x}, \boldsymbol{p}, t)}{\partial t} + \boldsymbol{p} \cdot \nabla_x n(\boldsymbol{x}, \boldsymbol{p}, t) = c\int d\boldsymbol{\eta} \sigma(\boldsymbol{p}, \boldsymbol{\eta}) \left( n(\boldsymbol{x}, \boldsymbol{\eta}, t) - n(\boldsymbol{x}, \boldsymbol{p}, t) \right)  ,
\end{align*}
where $n(\bx, \bp, t)$ is the Wigner distribution at position $\bx$, wave vector $\bp$ and time $t$, and $c \sigma(\boldsymbol{p}_1, \boldsymbol{p}_2)$ is the scattering cross section.

Interestingly, the evolution of the Wigner distribution predicted from the wave kinetic equation is an irreversible relaxation process. For this problem, a Lyapunov function, $S = \int d\boldsymbol{x} d\boldsymbol{p} \log n(\boldsymbol{x}, \boldsymbol{p})$, monotonically increases with time, even though the fundamental equation of motion possesses a time-reversal symmetry. This old irreversibility paradox has been recently revisited for the kinetic theory of particles \cite{bouchet2020boltzmann}, using dynamical large deviation principles, in the case of the Boltzmann equation. It turns out that the dynamical large deviation principle that quantifies the probability for the evolution of any trajectory has a time reversal symmetry. The kinetic equation corresponds to the most probable path of the system, while the probability of a path and its time-reversed path is related through detailed balance, a manifestation of time reversal symmetry for the mesoscopic stochastic process. This gives an extremely simple and enlightening explication of the irreversibility paradox. The main result of the present paper is a large deviation principle for the Schr\"odinger equation in a weak random potential, which has also a time reversal symmetry. This gives a new very clear explanation of the time reversal paradox.

The purpose of this paper is to formulate a path large deviation principle for wave scattering by random disorder in spatially inhomogeneous problems. In particular, to make the discussion as concise as possible, we restrict our attention to the simplest Schr\"odinger equation model. Our fundamental results are as follows. First, (i) for a small but finite $\mu$ we show that the probability that a path of local spectral density $\left\{ n^\mu(t) \right\}$ evolves at a vicinity of a prescribed specific path $\left\{ n (t) \right\}$ satisfies a large deviation principle:
\begin{align*}
& \mathbb{P} \left[ \left\{ n^\mu(t) = n(t) \right\}_{0 \leqslant t \leqslant t_f} \right] \\
\underset{\mu \to 0}{\asymp} & \exp \left( - \frac{1}{(2 \pi \mu)^d} \int_0^{t_f} dt \sup_\lambda \left\{ \int \lambda \dot{n} - \mathcal{H}[n, \lambda] \right\} \right) \mathbb{P}_0[n(0)],
\end{align*}
where $\mathcal{H}$ is the large deviation Hamiltonian that generally governs the stochastic fluctuations of macroscopic variables and  $\mathbb{P}_0[n(0)]$ is the probability of the initial condition $n^\mu(t=0)$. Obtaining the explicit expression for $\mathcal{H}$ from the microscopic equation is one of the main results of this paper. Next, (ii) we verify that the ordinary wave kinetic equation describes the path that minimizes the exponent of the probability functional. Then, (iii) we establish a large deviation principle for the microcanonical measure that defines the quasipotential of the mesoscopic stochastic process of the local spectral density. We analyze (iv) the property of the large deviation Hamiltonian, check its symmetries related to conservation laws and the time-reversal symmetry, and derive an expression of the detailed balance that connects the probabilities of a path and its time-reversed path. Finally (v) we study the diffusive limit when the scales of variation of the random potential are much larger than the typical wavelength of the waves. For this case we obtain a diffusive large deviation Hamiltonian, for which we check all the desired symmetries.

The paper is organized in the following order. In section \ref{sec:WKE}, we first set up the basic problem, introduce scaling and statistical assumptions, and derive the ordinary form of the wave kinetic equation. In section \ref{sec:LDT}, we derive the path large deviation principle for the temporal variations of the local spectral density. The approach has some analogy with that of \cite{GBE}, with a slightly different scaling assumption, and working with the Wigner distribution to describe the wave local spectral density. We then show that this Hamiltonian satisfies the expected properties. A remarkable point is that the quasipotential entering into the detailed balance relation is consistent with the one obtained from a direct computation of microcanonical ensemble, formulated in Appendix \ref{sec:MCE}. Section \ref{sec:perspective} proposes several possible extensions in future studies.

\section{Wave kinetic equation for a linear Schr\"odinger equation}\label{sec:WKE}

\subsection{Problem setup}
We consider the Schr\"odinger equation for a wave function $\psi^\ast (\boldsymbol{x}^\ast, t^\ast): \mathbb{R}^{d+1} \to \mathbb{C}$, where $t^\ast$ is time and $\bx^\ast$ is the position vector, and with potential $V^\ast(\boldsymbol{x}^\ast): \mathbb{R}^{d} \to \mathbb{R}$:
\begin{align} \label{eq:schrodinger}
i \frac{\partial \psi^\ast}{\partial t^\ast} = - \frac{D}{2} \nabla_{x^\ast}^2 \psi^\ast + V^\ast \psi^\ast .
\end{align}
We use an upper script $\ast$ to represent variables with physical dimensions. The physical parameter $D>0$  has dimension {$L^2 T^{-1}$. In absence of interaction with the potential, the free Schrödinger equation is the dynamics of linear waves with a dispersion relation $\omega(\bk^{\ast})= D \vert \bk^{\ast} \vert^{2} / 2$, where $\bk^{\ast}$ is a wave vector. A localised wave packet propagates at group velocity $\nabla_{k^\ast} \omega(\bk^{\ast}) = D\bk^{\ast}$.

The potential $V^\ast(\boldsymbol{x}^\ast)$ is assumed to be a spatially homogeneous random field with zero-average, $\mathbb{E}[V^\ast] = 0$, with its spectral density given by
\begin{align}
\Pi^\ast(\boldsymbol{k}^\ast) = \frac{1}{(2 \pi)^d} \int_{\mathbb{R}^d} d\boldsymbol{y}^\ast e^{- i \boldsymbol{k}^\ast \cdot \boldsymbol{y}^\ast} \mathbb{E} \left[ V^\ast \left(\boldsymbol{x}^\ast + \frac{\boldsymbol{y}^\ast}{2} \right) V^\ast \left(\boldsymbol{x}^\ast - \frac{\boldsymbol{y}^\ast}{2} \right) \right] .
\end{align}
For homogeneous fields, the two-point correlation function $\mathbb{E}\left[V^{\ast}( \boldsymbol{x}_{1}^{\ast}) V^{\ast}(\boldsymbol{x}_2^{\ast})\right]$ depends only on the point separation $\boldsymbol{x}_1^{\ast} - \boldsymbol{x}_2^{\ast}$. The spectral density of the potential is the Fourier transform of the two-point correlation function $\mathbb{E}\left[V^{\ast}( \boldsymbol{x}_{1}^{\ast}) V^{\ast}(\boldsymbol{x}_2^{\ast})\right]$ with respect to $\boldsymbol{x}_1^{\ast} - \boldsymbol{x}_2^{\ast}$ and thus contains the same information. Note that a prescription of higher order cumulants would be needed to fully characterize the potential distribution. As we will see in the following, the higher order statistics of the potential will not affect the dynamics of the spectral density of the waves in the kinetic regime. Hence, although we do not specify all the cumulants, the potential needs not to be Gaussian.

In this article, we assume that the spectral density $\Pi^{\ast}$ is concentrated around wave vectors $\lvert \boldsymbol{k}^{\ast} \rvert \sim 2\pi / \lambda$ where $\lambda$ is the typical wavelength. In real space, the wavelength $\lambda$ is interpreted as the typical correlation length of the potential. For such a potential, the Schr\"odinger dynamics Eq. (\ref{eq:schrodinger}) may display \yo{a number of} different regimes, depending on the order of magnitude of $\lambda$ compared to typical wavelengths in the initial condition of $\psi^\ast$. For instance, if the initial condition is made of waves with wavelengths much smaller than $\lambda$, the wave-potential interaction corresponds to random but smooth refraction that leads to diffusion in the macroscopic limit \cite{bal2010kinetic,kafiabad2019diffusion,dong2020frequency,boas2020directional}. We will see in section \ref{sec:diffusive_limit} that this diffusive limit can be recovered from the wave kinetic regime. On the other hand, if the initial waves have wavelength much greater that $\lambda$, one faces a homogenization problem that is not described by the wave kinetic equation \cite{guRandom2016, zhangHomogenization2014}. In the present paper, we will focus on an intermediate regime, when the initial condition is made of waves with typical wavelengths which are of order $\lambda$. We also make an assumption that the potential term is very small compared to the Laplacian term, by setting
\begin{align}
\epsilon \equiv \frac{\lambda^2 V_0}{D} \ll 1 ,
\end{align}
where $V_0$ is a constant scaling the potential, i.e., the potential spectrum is typically $\Pi^\ast \sim V_0^2 \lambda^d$.
}

Wave energy or wave action measures the local amplitude of the signal. We denote $\ell$ the typical scale for spatial variation of wave action, and call it the mesoscopic spatial scale. We introduce the second natural dimensionless parameter
\begin{align}
\mu \equiv \frac{\lambda}{\ell}.
\end{align}
The kinetic limit is the limit $\mu \ll 1$. In the context of wave kinetics, we are interested in the statistical behavior of the system at the mesoscopic scale, avoiding chasing rapid phase oscillations at scale $\lambda$.
Since the group velocity of a wave packet is $D\lvert \boldsymbol{k}^\ast \rvert \propto D\lambda^{-1}$, the migration time of a wave packet over a mesoscopic distance $\ell$ is $\lambda^2 / D\mu$. We call this time the mesoscopic time. Choosing such mesoscopic units naturally yields the following dimensionless coordinates
\begin{align}
\boldsymbol{x} \equiv \mu \frac{\boldsymbol{x}^\ast}{\lambda}, \quad \boldsymbol{p} \equiv \lambda \boldsymbol{k}^\ast, \quad t \equiv \mu \frac{Dt^\ast}{\lambda^2} ,
\end{align}
where the scaled wave vector is now represented by $\boldsymbol{p}$ in a customary way of quantum mechanics with $\mu$ corresponding to the Dirac constant. Physically, the square of the absolute value of the wave function, $\lvert \psi^\ast  \rvert^2$, represents the wave action density that is proportional to energy, momentum, or number of particles contained in a unit volume. Therefore, on the dimensional ground, the wave function should be dependent on the scaling parameters as
\begin{align*}
\psi^\mu (\boldsymbol{x}) = \frac{\lambda^{d/2} \psi^\ast (\boldsymbol{x}^\ast)}{\mu^{d/2}} .
\end{align*}
The potential and its spectrum are scaled as
\begin{gather*}
V^\mu(\boldsymbol{x}) = \frac{V^\ast(\boldsymbol{x}^\ast)}{V_0 }, \quad \Pi (\boldsymbol{p}) = \frac{\Pi^\ast (\boldsymbol{k}^\ast)}{V_0^2 \lambda^d},
\end{gather*}
such that
\begin{align}\label{eq:def:potential_spectrum}
\Pi (\boldsymbol{p}) = \frac{1}{(2 \pi \mu)^d} \int_{\mathbb{R}^d} d\boldsymbol{y} e^{- i \boldsymbol{p} \cdot \boldsymbol{y} / \mu} \mathbb{E} \left[ V^\mu \left( \boldsymbol{x} + \frac{\boldsymbol{y}}{2} \right) V^\mu \left( \boldsymbol{x} - \frac{\boldsymbol{y}}{2} \right) \right].
\end{align}
In the end, the governing equation (\ref{eq:schrodinger}) is rewritten in the non-dimensional form
\begin{align} \label{eq:schrodinger_scaled}
i \mu \frac{\partial \psi^\mu}{\partial t} = - \frac{\mu^2}{2} \nabla_x^2 \psi^\mu + \epsilon V^\mu \psi^\mu ,
\end{align}
which is the fundamental model of the present work.

For $\epsilon=0$, waves $e^{i(\boldsymbol{p} \cdot \boldsymbol{x} - \omega t) / \mu}$, with $\omega = \lvert \boldsymbol{p} \rvert^2 / 2$, are exact solutions of the equations, illustrating that the microscopic time scale and spatial scales are $t_m \sim \mathcal{O}(\mu)$ and  $\boldsymbol{x}_m \sim \mathcal{O}(\mu)$\yo{,} respectively. A wave packet $a(\boldsymbol{x},t)e^{i(\boldsymbol{p} \cdot \boldsymbol{x} - \omega t) / \mu}$ with modulation of its amplitude $a$ on spatial scales of order one (slow modulation compared to the microscopic scale), will actually see an evolution of $a$ on time scales of order one, according to the group velocity $\boldsymbol{p}$. In the limit of small $\epsilon$, the effect of the inhomogeneous potential term is very small on the microscopic time scale. Therefore, a wave packet propagates almost freely in a microscopic time scale. Since $\mathbb{E}[V^\mu] = 0$ has been assumed, effects of terms proportional to $\epsilon$ will vanish on average. Accumulation of the random potential effect will then give non zero contribution of order $\epsilon^2$. In order for this to be on the same order of magnitude as effects of free propagation on the wave action requires
  \begin{equation}
    \label{eq:def:kinetic_scaling}
    \epsilon = \sqrt{c\mu}
  \end{equation}
with $c>0$ a finite constant which accounts for the relative importance of the scattering interactions with respect to propagation. The constant $c$ is strictly speaking not needed, and it could be absorbed in the definition of $\Pi$, but it is useful for the physical discussion. The kinetic regime, or kinetic limit, is the joint limit $\mu \to 0$ with $\epsilon=\sqrt{c\mu}$, where $c$ is a fixed constant. Consequently, the pertinent equation for the kinetic scaling will be
\begin{align} \label{eq:schrodinger_scaled_c}
i \mu \frac{\partial \psi^\mu}{\partial t} = - \frac{\mu^2}{2} \nabla_x^2 \psi^\mu + \sqrt{c\mu} V^\mu \psi^\mu.
\end{align}
In some parts of the following sections, we will perform asymptotic expansions of the effect of the random potential by expanding (\ref{eq:schrodinger_scaled}) in power of $\epsilon$, while integrating out explicitly the wave propagation effects. For this reason, we often consider (\ref{eq:schrodinger_scaled}) instead of (\ref{eq:schrodinger_scaled_c}) for those technical parts.

\subsection{Local spectral density}
In the regime of wave kinetics, one is interested in the amount of wave action existing at each position and wave vector. This is provided by the (rescaled) Wigner distribution of the signal that is defined, following previous work on inhomogeneous wave kinetics \cite{ryzhik1996transport,bal2005kinetics}, as
\begin{align} \label{eq:Wigner_distribution_scaled}
n^\mu (\boldsymbol{x}, \boldsymbol{p}, t)  &  =  \frac{1}{(2 \pi \mu)^d} \int_{\mathbb{R}^d} d\boldsymbol{y} \, e^{- i \boldsymbol{p} \cdot \boldsymbol{y} / \mu} \psi^\mu \left(\boldsymbol{x} + \frac{\boldsymbol{y}}{2} \right) \psi^{\mu\dag} \left(\boldsymbol{x} - \frac{\boldsymbol{y}}{2} \right) \, ,
\end{align}
where we have denoted by $\psi^{\mu\dag}$ the complex conjugate of $\psi^{\mu}$.
The Wigner distribution function is the local spectral density of the wave action defined in both space of position and wave vector. Indeed, the integral of $n^\mu$ over wave-vector space coincides with the action density in position space, such that $\int_{\mathbb{R}^d} d\boldsymbol{p} \, n^\mu(\boldsymbol{x}, \boldsymbol{p}) = \lvert \psi^\mu(\boldsymbol{x}) \rvert^2$. A caution for the physical interpretation of $n^\mu$ is that it allows the existence of negative values, in contrast to wave action or energy that should be strictly non-negative. The negativity of $n^\mu$ can be eliminated by averaging it over a scale comparable to $\mu$ \citep{lee1995theory}.

In this study, we will take the asymptotically small limit of both $\mu$ and $\epsilon$. Statistically, the limit of $\mu \to 0$ is regarded as a kind of thermodynamic limit; since $\mu$ represents the typical correlation length of the wave signal, the total number of degrees of freedom increase as $\mu^{-d}$. In general, the thermodynamic limit makes sense when we specify the macroscopic or mesoscopic variables, e.g., temperature or pressure for gas molecules. In the present model, the potential spectrum $\Pi(\boldsymbol{p})$ is a mesoscopic control parameter that should not depend on $\mu$, and the local spectral density $n^\mu$ is a mesoscopic variable to be determined. In our formulation, a superscript $\mu$ is put on a variable that depends on the scaling parameter. It is worth keeping in mind that, although the spectral density function $\Pi(\boldsymbol{p})$ is fixed, the corresponding potential function $V^\mu(\boldsymbol{x})$ depends on $\mu$ because its structure becomes finer and finer for $\mu \to 0$.

\subsection{Conservation properties}\label{sec:conservation}
We consider equation (\ref{eq:schrodinger_scaled}) on a spatial domain $\Gamma$, which is either $\mathbb{R}^d$ or a periodic domain $\mathbb{V}$. The norm $\int_{\Gamma} d\boldsymbol{x} \lvert \psi^\mu (\boldsymbol{x}, t) \rvert^2$ is conserved by the dynamics. If $\Gamma$ is $\mathbb{R}^d$, the norm can be either finite for localized solution, or infinite. A local conservation law always exists for $\lvert \psi^\mu (\boldsymbol{x}, t) \rvert^2$.

We now consider the other invariants, for equation (\ref{eq:schrodinger_scaled}), or the associated local conservation laws. If not specified, integrations for positions and wave vectors are always performed over $\Gamma$ and $\mathbb{R}^d$, respectively.
For a polynomial function $f: \mathbb{R} \to \mathbb{R}$, we define an operator $\mathcal{K}^{\epsilon, \mu}_f = f(- (\mu^2 / 2) \nabla^2_x + \epsilon V^\mu)$. Since the Schrödinger operator is self-adjoint, a straightforward computation shows that
\begin{align} \label{eq:Conservation_K_psi}
\left< f \right> \equiv \int d\boldsymbol{x} \psi^{\mu \dag} (\boldsymbol{x}, t) \mathcal{K}^{\epsilon, \mu}_f \psi^\mu (\boldsymbol{x}, t)
\end{align}
is independent of the time $t$ when $\psi^{\mu}$ is a solution of Eq. \eqref{eq:schrodinger_scaled}. Equation \eqref{eq:Conservation_K_psi} is rewritten in terms of the local spectral density as
\begin{align}
\left< f \right> = \int d\boldsymbol{x} d\boldsymbol{p} K^{\epsilon, \mu}_f(\boldsymbol{x}, \boldsymbol{p}) n^\mu (\boldsymbol{x}, \boldsymbol{p}, t) ,
\end{align}
where
\begin{align*}
& K^{\epsilon, \mu}_f(\boldsymbol{x}, \boldsymbol{p})  = \int_{\mathbb{R}^d} d\boldsymbol{y} e^{- i \boldsymbol{p} \cdot \boldsymbol{y} / \mu} \hat{K}^{\epsilon, \mu}_f \left( \boldsymbol{x} + \frac{\boldsymbol{y}}{2}, \boldsymbol{x} - \frac{\boldsymbol{y}}{2} \right) \\
\yo{\text{with}} \quad &
\hat{K}^{\epsilon, \mu}_f(\boldsymbol{x}_1, \boldsymbol{x}_2)  \equiv \mathcal{K}^{\epsilon, \mu}_f \delta(\boldsymbol{x}_1 - \boldsymbol{x}_2)
\end{align*}
is the Weyl symbol of the operator $\mathcal{K}^{\epsilon, \mu}_f$. In general, $K^{\epsilon, \mu}_f$ can be expanded as a power series of $\epsilon$ and $\mu$. The leading-order term is immediately obtained as $K^{0, 0}_f(\boldsymbol{x}, \boldsymbol{p}) = f \left( \lvert \boldsymbol{p} \rvert^2 / 2 \right)$. Therefore, for the small limit of $\epsilon$ and $\mu$, we may write the general invariant as
\begin{align} \label{eq:invariant_f}
\left< f \right> = \int d\boldsymbol{x} d\boldsymbol{p} \, f \left( \frac{\lvert \boldsymbol{p} \rvert^2}{2} \right) n^\mu \left( \boldsymbol{x}, \boldsymbol{p} \right) .
\end{align}

Since the choice of $f$ is arbitrary, the present system possesses an infinite number of invariant. This property is related to wave frequency conservation. For a scattering of wave action in spectral space by a time-independent potential, wave frequency does not change. Therefore, the amount of wave action with frequency less than $\omega$ remains constant. Setting $f(\sigma) = h(\omega - \sigma)$ in (\ref{eq:invariant_f}), where $h$ is the Heaviside function and $\omega \in \mathbb{R}^+,$ we define
\begin{align} \label{eq:Definition_A}
\mathcal{A}_\omega [n^\mu] \equiv \int d\boldsymbol{x} d\boldsymbol{p} \, h \left(\omega - \frac{\lvert \boldsymbol{p} \rvert^2}{2} \right) n^\mu (\boldsymbol{x}, \boldsymbol{p}).
\end{align}
Conservation of $\left< f \right>$ for an arbitrary $f$ is equivalent to the conservation of $\mathcal{A}_\omega$ for an arbitrary $\omega \in \mathbb{R}^+$, as
$\left< f \right> = \int_{\mathbb{R}^+} d\omega \, \frac{d \mathcal{A}_\omega}{d \omega} f(\omega)$.

\subsection{Wave kinetic equation}\label{subsec:kinetic_equation}
In this subsection, we shall derive the classical form of the wave kinetic equation. A common derivation of the wave kinetic equation starts from a closed equation on the Wigner distribution and performs a multiple time-scale expansion, see e.g.~\cite{ryzhik1996transport}. We rather adopt here a perturbative approach in $\epsilon$ from the Schrödinger equation \eqref{eq:schrodinger_scaled}. This approach is also classical in the wave turbulence literature \cite{nazarenko2011, eyink2012a, GBE}, will appear helpful in the derivation of the dynamical large deviation theory in section \ref{sec:LDT}, and has the advantage to generalize easily to the case of the kinetic theory of non-linear waves with 3-wave interactions.

The first step is to express the solution of the Schrödinger equation  \eqref{eq:schrodinger_scaled} using an expansion in power of $\epsilon$, such that $\psi^\mu = \psi^\mu_0 + \epsilon \psi^\mu_1 + \epsilon^2 \psi^\mu_2 + \mathcal{O}(\epsilon^{3})$. Inserting this expansion to (\ref{eq:schrodinger_scaled}), we derive the series of equations:
\begin{align*}
i \mu \frac{\partial \psi^\mu_0}{\partial t} & = - \frac{\mu^2}{2} \nabla_x^2 \psi_0^\mu \\
i \mu \frac{\partial \psi^\mu_1}{\partial t} & = - \frac{\mu^2}{2} \nabla_x^2 \psi_1^\mu + V^\mu \psi_0^\mu \\
i \mu \frac{\partial \psi^\mu_2}{\partial t} & = - \frac{\mu^2}{2} \nabla_x^2 \psi_2^\mu + V^\mu \psi_1^\mu \\
& \vdots .
\end{align*}
We consider this expansion for any given initial condition $\psi^\mu (\boldsymbol{x}, 0) = \psi^{\mu,0}(\boldsymbol{x})$. We assume that the initial condition does not depend on $\epsilon$, i.e., $\psi^\mu_0(\boldsymbol{x}, 0) = \psi^{\mu,0}(\boldsymbol{x})$ and $\psi^\mu_j(\boldsymbol{x}, 0) = 0$ for $j \geq 1$. To write down the solution, we introduce the propagator $G^\mu(\boldsymbol{x}, t)$ such that
\begin{align} \label{eq:propagator_1}
\left( i \mu \frac{\partial}{\partial t} + \frac{\mu^2}{2} \nabla_x^2 \right) G^\mu = i \mu \delta(\boldsymbol{x}) \delta(t) ,
\end{align}
and $G^\mu = 0$ for $t < 0$. We then obtain
\begin{subequations} \label{eq:expansion_psi}
\begin{align}
\psi^\mu_0 (\boldsymbol{x}, t) & = \int_{\mathbb{R}^d} d\boldsymbol{\xi} G^\mu (\boldsymbol{x} - \boldsymbol{\xi}, t) \psi^\mu (\boldsymbol{\xi}, 0) \\
\psi^\mu_1 (\boldsymbol{x}, t) & = \frac{1}{i \mu} \int^t_0 d\tau \int_{\mathbb{R}^d} d\boldsymbol{\xi} G^\mu (\boldsymbol{x} - \boldsymbol{\xi}, t - \tau) V^\mu(\boldsymbol{\xi}) \psi^\mu_0 (\boldsymbol{\xi}, \tau) \nonumber \\
& = \frac{1}{i \mu} \int^t_0 d\tau \int_{\mathbb{R}^{2d}} d\boldsymbol{\xi}_{12} G^\mu (\boldsymbol{x} - \boldsymbol{\xi}_1, t - \tau) V^\mu(\boldsymbol{\xi}_1) G^\mu (\boldsymbol{\xi}_1 - \boldsymbol{\xi}_2, \tau) \psi^\mu (\boldsymbol{\xi}_2, 0) \\
\psi^\mu_2 (\boldsymbol{x}, t) & = \frac{1}{i \mu} \int^t_0 d\tau \int_{\mathbb{R}^d} d\boldsymbol{\xi} G^\mu (\boldsymbol{x} - \boldsymbol{\xi}, t - \tau) V^\mu(\boldsymbol{\xi}) \psi^\mu_1 (\boldsymbol{\xi}, \tau) \nonumber \\
& = \frac{1}{- \mu^2} \int^t_0 d\tau_1 \int^{\tau_1}_0 d\tau_2 \int_{\mathbb{R}^{3d}} d\boldsymbol{\xi}_{123} G^\mu (\boldsymbol{x} - \boldsymbol{\xi}_1, t - \tau_1) V^\mu(\boldsymbol{\xi}_1) \nonumber \\
& \times G^\mu (\boldsymbol{\xi}_1 - \boldsymbol{\xi}_2, \tau_1 - \tau_2) V^\mu(\boldsymbol{\xi}_2) G^\mu (\boldsymbol{\xi}_2 - \boldsymbol{\xi}_3, \tau_2) \psi^\mu (\boldsymbol{\xi}_3, 0) .
\end{align}
\end{subequations}
The equation for the propagator (\ref{eq:propagator_1}) is analytically solved as
\begin{align} \label{eq:propagator}
G^\mu (\boldsymbol{x}, t) = \frac{h(t)}{(2 \pi \mu)^d} \int d\boldsymbol{p} e^{- i \lvert \boldsymbol{p} \rvert^2 t / 2 \mu} e^{i\boldsymbol{p} \cdot \boldsymbol{x} / \mu} ,
\end{align}
where $h(t)$ is again the Heaviside function. Although the wave vector integration of this expression can be carried out, we keep this form because it is convenient for later computations.

Importantly, the perturbation solution will be valid for not too large $t$. For longer times, the higher order terms will not be small compared to $\psi^\mu_0$ even though $\epsilon$ is small. As will be clear in the following discussion, we will need the perturbative solution to be valid up to $\mu \ll t \ll 1$; an intermediate range between the microscopic and mesoscopic time scales.

Based on the perturbation solution of $\psi^\mu$ derived above, we shall consider the evolution of the local spectral density $n^\mu$. To simplify the computation, we introduce the Wigner transform of two functions, $f(\boldsymbol{x})$ and $g(\boldsymbol{x})$, as
\begin{align} \label{eq:Wigner_transform}
w^\mu (f, g) (\boldsymbol{x}, \boldsymbol{p}) \equiv \frac{1}{(2 \pi \mu)^d} \int_{\mathbb{R}^d} d\boldsymbol{y} \, f \left( \boldsymbol{x} + \frac{\boldsymbol{y}}{2} \right) g^\dag \left( \boldsymbol{x} - \frac{\boldsymbol{y}}{2} \right) e^{- i \boldsymbol{p} \cdot \boldsymbol{y} / \mu} .
\end{align}
Its inverse is
\begin{align}
f (\boldsymbol{x}_1) g^\dag (\boldsymbol{x}_2) = \int d\boldsymbol{p} \, w^\mu \left( \frac{\boldsymbol{x}_1 + \boldsymbol{x}_2}{2}, \boldsymbol{p} \right) e^{i \boldsymbol{p} \cdot (\boldsymbol{x}_1 - \boldsymbol{x}_2) / \mu} .
\end{align}
Basically, the local spectral density is the Wigner transform of identical wave functions,
\begin{align} \label{eq:Wigner_short}
n^\mu (\boldsymbol{x}, \boldsymbol{p}, t) = w^\mu (\psi^\mu, \psi^\mu) .
\end{align}
Inserting $\psi^\mu = \psi^\mu_0 + \epsilon \psi^\mu_1 + \epsilon^2 \psi^\mu_2 + \ldots$ to (\ref{eq:Wigner_short}), and using (\ref{eq:expansion_psi}), we obtain an expansion of the local spectral density $n^\mu$ in terms of $\epsilon$ as
\begin{align}
n^\mu (\boldsymbol{x}, \boldsymbol{p}, t) & = w^\mu (\psi^\mu_0, \psi^\mu_0) + \epsilon w^\mu (\psi^\mu_1, \psi^\mu_0) + \epsilon w^\mu (\psi^\mu_0, \psi^\mu_1) \nonumber \\
& + \epsilon^2 w^\mu (\psi^\mu_1, \psi^\mu_1) + \epsilon^2 w^\mu (\psi^\mu_2, \psi^\mu_0) + \epsilon^2 w^\mu (\psi^\mu_0, \psi^\mu_2) + \mathcal{O} (\epsilon^3) .
\label{n_mu}
\end{align}
The first term on the right-hand side is easily computed as
\begin{align} \label{eq:wave_migration}
w^\mu (\psi^\mu_0, \psi^\mu_0) & = n^\mu (\boldsymbol{x} - \boldsymbol{p}t, \boldsymbol{p}, 0) .
\end{align}
This expression means that, in absence of the potential, the propagation of free waves transports the spatial distribution of wave action density at the group velocity $\boldsymbol{p}$. It is notable that Eq. (\ref{eq:wave_migration}) is valid without taking an ensemble average or an asymptotic limit.

We shall consider the expectation of (\ref{n_mu}) with respect to the realization of the random potential $V^\mu$. Since $\mathbb{E}[V^\mu] = 0$ has been assumed, terms proportional to $\epsilon$ vanish. The dominant contribution from the random potential to the variations in local spectral density comes from terms of order $\epsilon^2$. Direct computations, performed in Appendix \ref{sec:scatter}, yield the expectation values of perturbation terms at the leading-order, (\ref{eq:term1-1}), (\ref{eq:term1-2}) and (\ref{eq:term1-3}). Consequently, we obtain
\begin{align*}
& \mathbb{E} \left[ \epsilon^2 w^\mu (\psi^\mu_1, \psi^\mu_1) + \epsilon^2 w^\mu (\psi^\mu_2, \psi^\mu_0) + \epsilon^2 w^\mu (\psi^\mu_0, \psi^\mu_2) \right] \\
= & \frac{\epsilon^2 t}{\mu} \int d\boldsymbol{\eta} \sigma(\boldsymbol{p}, \boldsymbol{\eta}) \left( n^\mu(\boldsymbol{x}, \boldsymbol{\eta}, 0) - n^\mu(\boldsymbol{x}, \boldsymbol{p}, 0) \right) + {\rm h.o.t.}
\end{align*}
with
\begin{align}\label{eq:def:cross_section}
\sigma(\boldsymbol{p}_1, \boldsymbol{p}_2) \equiv 2 \pi \Pi(\boldsymbol{p}_1 - \boldsymbol{p}_2) \delta \left( \frac{\lvert \boldsymbol{p}_1 \rvert^2}{2} - \frac{\lvert \boldsymbol{p}_2 \rvert^2}{2} \right) .
\end{align}
Here, h.o.t. represents the higher-order terms in the expansion that are negligible in the asymptotic limit. Setting $\epsilon = \sqrt{c\mu}$, we have
\begin{align*}
& \lim_{\mu \to 0} \frac{\mathbb{E}[n^\mu(\boldsymbol{x}, \boldsymbol{p}, t)] - n^\mu(\boldsymbol{x} - \boldsymbol{p} t, \boldsymbol{p}, 0)}{t} \\
= & \lim_{\mu \to 0} c \int d\boldsymbol{\eta} \sigma (\boldsymbol{p}, \boldsymbol{\eta}) \left( n^\mu(\boldsymbol{x}, \boldsymbol{\eta}, 0) - n^\mu(\boldsymbol{x}, \boldsymbol{p}, 0) \right) .
\end{align*}
Taking the limit $t \to 0$, we obtain a differential equation for the ensemble average of the local spectral density, $\lim_{\mu \to 0} \mathbb{E}[n^\mu] = n$:
\begin{align*}
\left. \frac{\partial n (\boldsymbol{x}, \boldsymbol{p}, t)}{\partial t} \right\vert_{t=0} + \boldsymbol{p} \cdot \nabla_x n (\boldsymbol{x}, \boldsymbol{p}, 0) = c \int d\boldsymbol{\eta} \sigma (\boldsymbol{p, \boldsymbol{\eta}}) \left( n(\boldsymbol{x}, \boldsymbol{\eta}, 0) - n(\boldsymbol{x}, \boldsymbol{p}, 0) \right) .
\end{align*}
At this stage, this expression is valid only at the initial time, $t = 0$. One cannot extend it to $t>0$ because the wave function $\psi^\mu$ is \emph{a priori} correlated with the potential field $V^\mu$. However, we shall argue that the correlation between $\psi^{\mu}$ and $V^{\mu}$ remains always weak at any time. Indeed, in the present problem, it is assumed that the significant modification of the wave field $\psi^\mu$ by scattering on the potential occurs at a time scale of wave propagation over a mesoscopic distance. The typical correlation length of the random potential $V^\mu$ is much shorter than this mesoscopic scale by a factor of $\mu$. Therefore, even though interferences of the field with the potential produce slight correlations, the free propagation of the field makes the correlation vanishes. This situation resembles the loss of memory for particle collision in dilute gas\textemdash the molecular chaos hypothesis. The present weak-correlation assumption allows us to regard the temporal evolution of $n^\mu$ as a Markovian process such that the wave kinetic equation would be valid any time as far as $\mu$ is sufficiently small. \yo{We note that this explanation does not apply to linear systems in cases when Anderson's localization occurs \cite{lagendijkFifty2009,shengIntroduction2006}. In particular for $d=1$ cases, it is known that localization is intense so that the kinetic description is not appropriate \cite{bal2005kinetics,bal2010kinetic}. Interestingly, the presence of weak nonlinearities may destroy strong localization and one can recover somehow a modified kinetic regime \cite{nazarenko2019wave} that is beyond the scope of the present paper. For $d>1$, obtaining the general criteria of localization for (linear) systems is much more challenging and will not be discussed in this work. We thus assume in this article a dimension $d>1$ and regimes where no localization phenomena do occur. 
}


Once the Markovian hypothesis is accepted, we may iterate the reasoning expounded above for $t=0$ in order to reach any time $t>0$. We eventually obtain the equation
\begin{align} \label{eq:WKE}
\frac{\partial n (\boldsymbol{x}, \boldsymbol{p}, t)}{\partial t} + \boldsymbol{p} \cdot \nabla_x n (\boldsymbol{x}, \boldsymbol{p}, t) = c \int d\boldsymbol{\eta} \sigma (\boldsymbol{p}, \boldsymbol{\eta}) \left( n(\boldsymbol{x}, \boldsymbol{\eta}, t) - n(\boldsymbol{x}, \boldsymbol{p}, t) \right) ,
\end{align}
\yo{which} is the ordinary form of the wave kinetic equation. Since the group velocity of a free wave is now $\boldsymbol{p}$, the second term on the left-hand side is understood as the motion of the Wigner distribution at the group velocity due to the free dynamics. The right-hand side represents wave scattering in wave-vector space that occurs at microscopic scale. The function $c\sigma(\boldsymbol{p}_1, \boldsymbol{p}_2)$ is the scattering cross section determining the rate of wave action converted from wave vector $\boldsymbol{p}_1$ to $\boldsymbol{p}_2$ per unit time.

The wave kinetic equation inherits the conservation property of the original Schr\"odinger equation, namely that $d \mathcal{A}_\omega[n] / dt = 0$. However, the wave kinetic equation differs from the microscopic Schr\"{o}dinger dynamics (\ref{eq:schrodinger_scaled}) since the former appears time-irreversible whereas the latter is time reversible. Indeed, for a prescribed potential field $V^\mu(\boldsymbol{x})$, let $\psi^{\mu}(\boldsymbol{x}, t)$ be a solution of (\ref{eq:schrodinger_scaled}). Its time-reverse counterpart is defined as $\psi^{\mu}_R(\boldsymbol{x}, t) = \psi^\dag(\boldsymbol{x}, - t)$. As a consequence, the time-reversed local spectral density reads as $n_R^\mu (\boldsymbol{x}, \boldsymbol{p}, t) = n^\mu (\boldsymbol{x}, - \boldsymbol{p}, - t)$. Change in sign of $\boldsymbol{p}$ is a natural outcome because the wave group velocities of the forward and the reverse paths should be opposite. If $\psi^{\mu}(\boldsymbol{x}, t)$ is a solution of the Schr\"odinger equation, $\psi^{\mu}_R(\boldsymbol{x}, t)$ is also a solution. This is the time-reversal symmetry.  By contrast, the wave kinetic equation violates this symmetry: if $n$ is a solution, the time-reversed $n_R$ is not a solution of the wave kinetic equation, unless both $n$ and $n_R$ are an identical stationary state (see Eq. (\ref{eq:stationary_WKE}) defined below). This irreversibility paradox is reminiscent to the one raised by Boltzmann for the case of dilute gases. As explained in \cite{bouchet2020boltzmann} for the case of the Boltzmann equation, we will see in the next section that one can recover time-reversibility for the kinetic theory at the large deviation level, as a time-reversibility for the stochastic process of the local spectral density. Time-irreversibility arises because the wave kinetic equation describes the evolution of the \emph{average} $n=\mathbb{E}[n^{\mu}]$ only, or equivalently in this case the most probable evolution only. Other paths, including time-reversed paths, are possible: they are just extremely unlikely.

Before moving toward the large deviation theory of wave kinetics, we can remark that time-irreversibility can be also quantified by introducing a Lyapunov function
\begin{align}\label{eq:quasipotential-unnorm}
S \equiv \int d\boldsymbol{x} d\boldsymbol{p} \log n(\boldsymbol{x}, \boldsymbol{p}) .
\end{align}
Following solutions of the wave kinetic equation, $S$ increases monotonically with time, $dS / dt \geq 0$. If the spatial domain $\Gamma$ is finite, $S$ achieves the maximum when the spectral density $n(\boldsymbol{x}, \boldsymbol{p})$ is homogeneous in $\boldsymbol{x}$ and isotropic in $\boldsymbol{p}$. We write this homogeneous distribution of the spectral density under the constraint of $\mathcal{A}_\omega[n] = A(\omega)$ as
\begin{align} \label{eq:stationary_WKE}
n^A_{h}(\boldsymbol{p}) = \dfrac{A'(\lvert \boldsymbol{p} \rvert^2 / 2)}{\int d\boldsymbol{x} d\boldsymbol{\boldsymbol{\eta}} \delta(\lvert \boldsymbol{p} \rvert^2 / 2 - \lvert \boldsymbol{\eta} \rvert^2 / 2)} ,
\end{align}
where the denominator is introduced for a normalization purpose, and $A' = dA / d\omega$. It is obvious that $n(\boldsymbol{x}, \boldsymbol{p}) = n^A_h( \boldsymbol{p} )$ is a stationary solution of the wave kinetic equation. We note that the Lyapunov function $S$ is actually the microcanonical entropy of the macrostate specified by $\left( n^{\mu}, \mathcal{A}_{\omega} \right) =(n, A)$ for the Schr\"odinger equation as we discuss in Appendix~\ref{sec:MCE}, and is related to the quasipotential that appears in the discussion of the large deviation theory, as we discuss in section \ref{sec:quasipotential}.

\section{Large deviation formulation for random wave scattering} \label{sec:LDT}
In the previous section, we derived the wave kinetic equation as an equation for the ensemble average $\mathbb{E}\left[n^\mu \right]$ of the empirical local spectral density, with respect to the probability measure of the random potential, in the limit $\mu \rightarrow 0$. This kinetic limit can be understood as a law of large number: $\lim_{\mu \to 0} n^\mu = \mathbb{E}\left[n^\mu \right]$, where the limit has to be understood in a weak sense (for instance, the limit holds when both $n^\mu$ and $\mathbb{E}\left[n^\mu \right]$ are integrated over any subset $\mathbb{U} \subset \Gamma \times \mathbb{R}^d$).

In statistical mechanics, it is quite common that an empirical macroscopic quantity converges to its ensemble average \yo{in} the large limit of the number of elements. The deviation from the average is often exponentially small and evaluated asymptotically by a large deviation principle. In this section, our aim is to generalize the law of large number for the kinetic theory and to compute the probability to observe any possible fluctuations for $n^\mu$, as a large deviation principle, in the limit $\mu \rightarrow 0$. Such fluctuations are expected to be characterized by a large deviation parameter proportional to $\mu^{d}$, where $d$ is the space dimension, because the number of statistically independent degrees of freedom is of order $\mu^{-d}$.

\subsection{Large deviation Hamiltonian}
We define the Newton ratio as the time increment for the local spectral density: $\Delta n^\mu/\Delta t = (n^\mu(\Delta t) - n^\mu(0)) / \Delta t$. We regard the temporal variations in $n^\mu$ as a stochastic process, and look for the probability to observe a value of the Newton ratio, conditioned on the value of the local spectral density at time 0: $n^\mu(0) = n$. Our aim is to justify that it satisfies
\begin{align}
- \lim_{\Delta t \to 0}  \lim_{\substack{\mu \to 0 \\ \epsilon = \sqrt{c\mu}}} \frac{(2 \pi \mu)^d}{\Delta t} \log \mathbb{P}\left[ \tfrac{\Delta n^\mu}{\Delta t} = \dot{n} \vert n^\mu = n \right] = \mathcal{L} \left[ n, \dot{n} \right],
\label{Lagrangian-definition}
\end{align}
and to derive an explicit formula for the Lagrangian $\mathcal{L}$. The limit $\lim_{\substack{\mu \to 0 \\ \epsilon = \sqrt{c\mu}}}$ corresponds to the kinetic limit $\mu \to 0$ where one has fixed $\epsilon =\sqrt{c \mu}$. Here, $\mathcal{L}[n, \dot{n}]$ is the rate function of the probability of the Newton ratio $\mathbb{P}\left[ \dot{n} \vert n \right]$.

Through the fast microscopic dynamics, the memory of the initial condition of the phases of $\psi^\mu$ are expected to be lost after some times, meaning that two-times, or multi-times, correlation functions are expected to decay with the time differences of two or several phase observables. Such a mixing property is expected to be due to the conjunction of phase mixing (oscillating integrals and the Riemann--Lebesgue lemma), spatial transport and dispersion, and the effect of the random potential. Because the natural timescale for phase dynamics is the microscopic timescale, one might expect a typical mixing time to be much smaller than the kinetic time scale and to decay to zero in the kinetic limit. This mixing property is required to justify a Markovian behavior and to propagate local in time results, like the Lagrangian (\ref{Lagrangian-definition}), in order to describe the dynamics at finite times. In none of the existing classical kinetic theories, neither physicists nor mathematicians have been so far able to justify or prove the requested mixing properties, in order to justify the long time validity of kinetic theories or their probabilistic large deviation generalizations. This is the main reason why mathematical results, for instance the celebrated Lanford's result for the Boltzmann equation \cite{lanfordTime1975a}, or its generalizations \cite{gallagher2013newton}, are usually valid only for a fraction of the kinetic time. As it is customary in theoretical physics, we will assume the validity of such a mixing property in the following.

We now assume the natural mixing hypothesis and the related Markov behavior of the stochastic process. As a consequence, the evolution of $n^\mu$ does not depend on its previous state. Then the path probability for the stochastic  process is Markovian, and the probability of a path of $n^\mu$ for a finite time interval, $0 \leqslant t \leqslant t_f$, can be derived from the local in time Lagrangian (\ref{Lagrangian-definition}). The path probability conditioned on the local spectral density at the initial time $n^\mu(0) = n(0)$ is then
\begin{align} \label{eq:LDT_Lagrangian}
\mathbb{P}_{n(0)} \left[ \left\{ n^\mu(t) = n(t) \right\}_{0 \leqslant t \leqslant t_f} \right] \underset{\mu \to 0}{\asymp} \exp \left( - \frac{1}{(2 \pi \mu)^d} \int_0^{t_f} dt \mathcal{L}[n, \dot{n}] \right) .
\end{align}
This expression is analogous to the path-integral formulation for the probability density in quantum theory. Note that the initial condition $n^{\mu}(0)=n(0)$ is fixed here, but one can easily consider a set of initial conditions complemented by an initial probability density $\mathbb{P}_0$. In such a case, the path probability of a trajectory $\{n(t)\}_{0\leqslant t \leqslant t_f}$ reads as
\begin{equation}
  \label{eq:LDT_set_initial_condition}
  \mathbb{P} \left[ \left\{ n^\mu(t) = n(t) \right\}_{0 \leqslant t \leqslant t_f} \right] \underset{\mu \to 0}{\asymp} \exp \left( - \frac{1}{(2 \pi \mu)^d} \int_0^{t_f} dt \mathcal{L}[n, \dot{n}] \right) \mathbb{P}_0[n(0)] .
\end{equation}

To compute the Lagrangian (\ref{Lagrangian-definition}), we will use the G\"artner-Ellis theorem that connects the rate function $\mathcal{L}$ to the cumulant-generating function, or the Hamiltonian $\mathcal{H}$ defined by
\begin{align} \label{eq:Hamiltonian_definition}
\mathcal{H}[n, \lambda] & = \lim_{ \Delta t \to 0} \lim_{\substack{\mu \to 0 \\ \epsilon = \sqrt{c\mu}}} \frac{(2 \pi \mu)^d}{\Delta t} \nonumber \\
& \times \log \mathbb{E} \left[ \exp \left( \frac{\int d\boldsymbol{x} d\boldsymbol{p} \lambda(\boldsymbol{x}, \boldsymbol{p}) (n^\mu(\boldsymbol{x}, \boldsymbol{p},\Delta t) - n(\boldsymbol{x}, \boldsymbol{p}))}{(2 \pi \mu)^d} \right) \right] ,
\end{align}
through the Legendre-Fenchel transform,
\begin{align} \label{eq:Lagrangian_Hamiltonian}
\mathcal{L} \left[ n, \dot{n} \right] \equiv \sup_\lambda \left\{ \int d\boldsymbol{x} d\boldsymbol{p} \lambda (\boldsymbol{x}, \boldsymbol{p}) \dot{n} (\boldsymbol{x}, \boldsymbol{p}) - \mathcal{H}[n, \lambda] \right\} .
\end{align}
Our aim here is to derive the specific form of $\mathcal{H}$ directly from the perturbation solutions of the original Schr\"odinger equation \eqref{eq:schrodinger_scaled}.

Following \cite{GBE}, we first compute the moment-generating function for the increment of $n^\mu$,
\begin{align} \label{eq:moment-generating}
\mathcal{Z}^\mu[n, \lambda; \Delta t] \equiv \mathbb{E} \left[ \exp \left( \frac{\int d\boldsymbol{x} d\boldsymbol{p} \lambda(\boldsymbol{x}, \boldsymbol{p}) (n^\mu(\boldsymbol{x}, \boldsymbol{p},\Delta t) - n(\boldsymbol{x}, \boldsymbol{p}))}{(2 \pi \mu)^d} \right) \right] .
\end{align}
We insert the expansion of $n^\mu(\boldsymbol{x}, \boldsymbol{p}, t)$ in terms of $\epsilon$ into (\ref{eq:moment-generating}). Since at the leading order $w^\mu (\psi^\mu_0, \psi^\mu_0)$ is statistically independent of $V^\mu$ by the weak correlation hypothesis described before, it is possible to decompose $\mathcal{Z}^\mu$ into two parts as
\begin{subequations}
\begin{align}
\mathcal{Z}^\mu & = \mathcal{Z}^\mu_0 \mathcal{Z}^\mu_\epsilon \\
\mathcal{Z}^\mu_0 & \equiv \exp \left( \frac{1}{(2 \pi \mu)^d} \int d\boldsymbol{x} d\boldsymbol{p} \lambda(\boldsymbol{x}, \boldsymbol{p}) \left( w^\mu (\psi^\mu_0, \psi^\mu_0) - n (\boldsymbol{x}, \boldsymbol{p}) \right) \right) \\
\mathcal{Z}^\mu_\epsilon & \equiv \mathbb{{E}} \left[ \exp \Biggl( \frac{1}{(2 \pi \mu)^d} \int d\boldsymbol{x} d\boldsymbol{p} \lambda(\boldsymbol{x}, \boldsymbol{p}) \left( n^\mu(\boldsymbol{x}, \boldsymbol{p}, \Delta  t) - w^\mu (\psi^\mu_0, \psi^\mu_0) \right) \Biggr) \right] .
\end{align}
\end{subequations}
Using (\ref{eq:wave_migration}), the first part, $\mathcal{Z}^\mu_0$, is immediately rewritten as
\begin{align}
\mathcal{Z}^\mu_0 = \exp \left( \frac{1}{(2 \pi \mu)^d} \int d\boldsymbol{x} d\boldsymbol{p} \lambda(\boldsymbol{x}, \boldsymbol{p}) \left( n(\boldsymbol{x} - \boldsymbol{p} \Delta t, \boldsymbol{p}, 0) - n (\boldsymbol{x}, \boldsymbol{p}) \right) \right) .
\end{align}
The second part is expanded in terms of $\epsilon$ to obtain
\begin{align}
\mathcal{Z}^\mu_\epsilon = 1 + \epsilon^2 \mathcal{Z}^\mu_2 + o(\epsilon^2) ,
\end{align}
where we have used $\mathbb{E}[V^\mu] = 0$. The Landau notation $o(\epsilon^{2})$ gathers all the terms that are negligible compared to $\mathcal{O}(\epsilon^{2})$ terms in the expansion. In the following, we shall discard the higher order terms of $o(\epsilon^2)$ and concentrate on computing $\mathcal{Z}^\mu_2$. Because we are considering the simultaneous limit of \yo{$\mu \to 0$ and $\epsilon=\sqrt{c\mu}$}, neglecting $o(\epsilon^2)$ terms cannot be justified {\it a priori}. This kind of problem commonly arises in wave kinetic theory \citep{GBE} but we expect $o(\epsilon^{2})$ to be negligible in the kinetic limit. The direct computation of $\mathcal{Z}^\mu_2$ yields
\begin{align}
\mathcal{Z}^\mu_2 & = \frac{1}{(2 \pi \mu)^d} \int d\boldsymbol{x} d\boldsymbol{p} \lambda(\boldsymbol{x}, \boldsymbol{p}) \mathbb{E} \left[ w^\mu(\psi^\mu_2, \psi^\mu_0) + w^\mu(\psi^\mu_0, \psi^\mu_2) + w^\mu(\psi^\mu_1, \psi^\mu_1) \right] \nonumber \\
& + \frac{1}{2 (2 \pi \mu)^{2d}} \int d\boldsymbol{x}_{12} d\boldsymbol{p}_{12} \lambda(\boldsymbol{x}_1, \boldsymbol{p}_1) \lambda(\boldsymbol{x}_2, \boldsymbol{p}_2) \nonumber \\
& \times \mathbb{E} \bigl[ (w^\mu(\psi^\mu_1, \psi^\mu_0)(\boldsymbol{x}_1, \boldsymbol{p}_1) + w^\mu(\psi^\mu_0, \psi^\mu_1)(\boldsymbol{x}_1, \boldsymbol{p}_1)) \nonumber \\
& \times (w^\mu(\psi^\mu_1, \psi^\mu_0)(\boldsymbol{x}_2, \boldsymbol{p}_2) + w^\mu(\psi^\mu_0, \psi^\mu_1)(\boldsymbol{x}_2, \boldsymbol{p}_2)) \bigr] .
\end{align}
From the computations in Appendix \ref{sec:scatter}, specifically (\ref{eq:term1-1})-(\ref{eq:term1-3}) and \eqref{eq:term2-1}-\eqref{eq:term2-4}, we obtain
\begin{align} \label{eq:Z2}
\mathcal{Z}^\mu_2 & = \frac{\Delta t}{\mu (2 \pi \mu)^d} \int d\boldsymbol{x} d\boldsymbol{p}_{12} \Bigl[ (\lambda(\boldsymbol{x}, \boldsymbol{p}_1) - \lambda(\boldsymbol{x}, \boldsymbol{p}_2)) \sigma(\boldsymbol{p}_1, \boldsymbol{p}_2) n(\boldsymbol{x}, \boldsymbol{p}_2) \nonumber \\
& + \frac{1}{2} (\lambda(\boldsymbol{x}, \boldsymbol{p}_1) - \lambda(\boldsymbol{x}, \boldsymbol{p}_2))^2 \sigma(\boldsymbol{p}_1, \boldsymbol{p}_2) n(\boldsymbol{x}, \boldsymbol{p}_1) n(\boldsymbol{x}, \boldsymbol{p}_2) \Bigr] + {\rm h.o.t.} .
\end{align}
Inserting $\mathcal{Z}^\mu = \mathcal{Z}^\mu_0 \left( 1 + \epsilon^2 \mathcal{Z}^\mu_2 \right)$ into $\mathcal{H} = \lim_{\Delta t \to 0} \lim_{\substack{\mu \to 0 \\ \epsilon = \sqrt{c\mu}}} ( (2 \pi \mu)^d / \Delta t ) \log \mathcal{Z}^\mu$ and again neglecting $o(\epsilon^2)$ terms, we obtain the Hamiltonian as
\begin{subequations}
\begin{align}
\mathcal{H}[n, \lambda] & = \mathcal{H}_F + \mathcal{H}_S  \label{eq:H}\\
\mathcal{H}_F & = - \int d\boldsymbol{x} d\boldsymbol{p} \lambda(\boldsymbol{x}, \boldsymbol{p}) \boldsymbol{p} \cdot \nabla_x n(\boldsymbol{x}, \boldsymbol{p}) \label{eq:transport_H}\\
\mathcal{H}_S & = c \int d\boldsymbol{x} d\boldsymbol{p}_{12} \Bigl[ (\lambda(\boldsymbol{x}, \boldsymbol{p}_1) - \lambda(\boldsymbol{x}, \boldsymbol{p}_2)) \sigma(\boldsymbol{p}_1, \boldsymbol{p}_2) n(\boldsymbol{x}, \boldsymbol{p}_2)  \nonumber \\
& + \frac{1}{2} (\lambda(\boldsymbol{x}, \boldsymbol{p}_1) - \lambda(\boldsymbol{x}, \boldsymbol{p}_2))^2 \sigma(\boldsymbol{p}_1, \boldsymbol{p}_2) n(\boldsymbol{x}, \boldsymbol{p}_1) n(\boldsymbol{x}, \boldsymbol{p}_2) \Bigr] . \label{eq:scattering_H}
\end{align}
\end{subequations}
We have separated the Hamiltonian into two parts, $\mathcal{H}_F$ and $\mathcal{H}_S$. The first part, $\mathcal{H}_F$, represents the free wave propagation in position space and the second one, $\mathcal{H}_S$, the wave scattering in wave-vector space by the random potential.

\subsection{Properties of the large deviation Hamiltonian} \label{sec:property_Hamiltonian}

Once the specific form of the Hamiltonian is obtained, we can discuss the properties of the stochastic process governed by the path-integral formula (\ref{eq:LDT_Lagrangian}) and (\ref{eq:Lagrangian_Hamiltonian}). Formulations in the paper \cite{bouchet2020boltzmann} are simple and informative, and we summarize several important properties of dynamical large deviation theory in Appendix \ref{sec:fluctuation}. We now check the classical expected properties of $\mathcal{H}$.

\subsubsection{Weak noise Langevin dynamics and wave kinetic equation}
The first important property of the large deviation Hamiltonian $\mathcal{H}$ is that it is quadratic and convex with respect to the conjugated field $\lambda$. This means that the fluctuations of the infinitesimal current $\dot{n}d t$ are, locally in time, Gaussian. Reading the quadratic part of the Hamiltonian (\ref{eq:scattering_H}), we see that the local covariance of the local in time Gaussian fluctuations are given by the diffusion kernel
\begin{align} \label{eq:def:operator_Sigma}
\Sigma[n](\bx;\bp_1,\bp_2) & = -c \sigma(\bp_1,\bp_2) n(\bx, \bp_1) n(\bx, \bp_2) \nonumber \\
& + c \int d\boldsymbol{\eta} \sigma(\bp_1,\boldsymbol{\eta}) n(\bx, \bp_1)  n(\bx, \boldsymbol{\eta}) \delta(\bp_1 -\bp_2).
\end{align}
As a consequence, in the kinetic limit $\mu\ll 1$, the dynamics of the local spectral density in the kinetic regime corresponds to a weak noise Langevin dynamics \cite{freidlinRandom2012,grahamMacroscopic1989}
\begin{align} \label{eq:weak_langevin_dynamics}
\dot{n}(\bx, \bp, t) & = - \bp \cdot \nabla_x n + c \int d  \boldsymbol{\eta} \, \sigma(\bp, \boldsymbol{\eta})\left(n(\bx,  \boldsymbol{\eta},t) - n(\bx, \bp, t)\right) \nonumber \\
& + \sqrt{2} (2\pi\mu)^{d/2} \int d  \boldsymbol{\eta} \, \Sigma^{1/2}[n](\bx ; \bp,  \boldsymbol{\eta}) \xi( \boldsymbol{\eta},t) ,
\end{align}
\yo{where} $\xi(\bp ,t)$ is a white noise such that $\mathbb{E}\left[ \xi(\bp ,t)\xi(\boldsymbol{\eta} ,s) \right]= \delta(t-s)\delta(\bp - \boldsymbol{\eta})$, and $\Sigma^{1/2}[n]$ is defined as a square root of \yo{the} diffusion kernel, meaning that $\int d\boldsymbol{\eta} \Sigma^{1/2} (\bx, \bp_1, \boldsymbol{\eta}) \Sigma^{1/2} (\bx, \boldsymbol{\eta}, \bp_2) = \Sigma (\bx, \bp_1, \bp_2)$.

The second expected property is that the most probable path is the solution of the wave kinetic equation. This is easily checked by noticing that the most probable path, for which the action $\int_0^{t_f}\mathcal{L}[n, \dot{n}]dt =0$ vanishes, satisfies
\begin{align}
\frac{\partial n}{\partial t} = \left. \frac{\delta \mathcal{H}}{\delta \mathcal{\lambda}} \right\vert_{\lambda = 0}
\end{align}
(see Appendix \ref{app:sub:relaxation_path}), and is also the linear term in $\lambda$ of the Hamiltonian $\mathcal{H}$ and the drift term of the Langevin dynamics (\ref{eq:weak_langevin_dynamics}). Then, the path large deviation analysis confirms that the wave kinetic equation can be understood as a law of large number at the level of trajectories.

It is enlightening to rewrite the scattering part of the Hamiltonian $\mathcal{H}_S$ in terms of the diffusion kernel (\ref{eq:def:operator_Sigma}) as
  \begin{equation}
    \label{eq:scattering_Hamiltonian_DB}
    \mathcal{H}_S = \int d\bx d\bp_{12} \, \lambda(\bx, \bp_1) \Sigma[n](\bx ; \bp_1 ,\bp_2) \left( \frac{\delta S}{\delta n(\bx, \bp_2)} + \lambda(\bx,\bp_2) \right),
  \end{equation}
 where $S$ is the entropy (\ref{eq:quasipotential-unnorm}), and the Langevin equation (\ref{eq:weak_langevin_dynamics}) as
 \begin{align} \label{eq:weak_langevin_dynamics_entropy}
\dot{n}(\bx, \bp, t) + \bp \cdot \nabla_x n & =  \int d  \boldsymbol{\eta} \, \Sigma[n](\bx ; \bp,  \boldsymbol{\eta}) \frac{\delta S}{\delta n(\bx, \boldsymbol{\eta})} \nonumber \\
& + \sqrt{2} (2\pi\mu)^{d/2} \int d  \boldsymbol{\eta} \, \Sigma^{1/2}[n](\bx ; \bp,  \boldsymbol{\eta}) \xi( \boldsymbol{\eta},t).
\end{align}
This suggestive \yo{form} immediately \yo{emphasizes} that $S$ is a Lyapunov function, and that the dynamics has a detailed balance structure, as further explained in section \ref{sec:detailed_balance}.

\subsubsection{Conservation of the wave action distribution}

As we discussed in section \ref{sec:conservation}, in the original Schr\"odinger equation, the quantity $\mathcal{A}_\omega[n^\mu] = \int d\boldsymbol{x} d\boldsymbol{p} h(\omega - \lvert \boldsymbol{p} \rvert^2 / 2) n^\mu (\boldsymbol{x}, \boldsymbol{p})$ for any $\omega \in \mathbb{R}^+$ is conserved. This conservation property has to be also verified at the large deviation level, meaning that any trajectory $\{n(t)\}_{0\leqslant t \leqslant t_f}$ has to lie on the manifold $\mathcal{A}_\omega[n]=A(\omega)$. As explained in Appendix~\ref{sec:conservation_appendix}, this is equivalent to the Hamiltonian symmetry
\begin{align}
\mathcal{H} \left[ n, \lambda + \alpha \frac{\delta \mathcal{A}_\omega}{\delta n} \right] = \mathcal{H} \left[ n, \lambda \right],
\end{align}
for an arbitrary $\alpha \in \mathbb{R}$, $n$ and $\lambda$. For the specific form of the Hamiltonian $\mathcal{H}=\mathcal{H}_F + \mathcal{H}_S$ \eqref{eq:H} with $\mathcal{H}_S$ written in the symmetric form (\ref{eq:scattering_Hamiltonian_DB}), one can directly check the above Hamiltonian symmetry boils down to the following property on the diffusive kernel (\ref{eq:def:operator_Sigma})
\begin{equation}
 \int d  \boldsymbol{\eta} \, \Sigma[n](\bx; \bp, \boldsymbol{\eta}) \frac{\delta \mathcal{A}_\omega}{\delta n(\bx, \boldsymbol{\eta})} = 0,
\end{equation}
for any $n$, $\bx$, $\bp$, and $\omega$.

\subsubsection{Stationary quasipotential} \label{sec:quasipotential}

It was shown in the previous section that the wave kinetic equation possesses an attractor which is a homogeneous distribution, $n_h$, with the prescribed constraints $\mathcal{A}_\omega[n] = A(\omega)$, for any $\omega$. We now consider the fluctuations of $n$ from $n_h$ at the large deviation level. More precisely, we define the equilibrium distribution of the stochastic process at a large deviation level:
\begin{align} \label{eq:stationary}
\mathbb{P}^\mu_{A,S}[\left\{ n^\mu = n \right\}] \underset{\mu\to0}{\asymp} \exp \left(-\frac{\mathcal{U}_A[n]}{(2 \pi \mu)^d} \right),
\end{align}
where $\mathbb{P}^\mu_{A,S}$ is the stationary probability measure of the microcanonical ensemble which is parameterized by a small constant $\mu$ as well as a function $A(\omega)$ specifying the action conservation constraint. The rate function $\mathcal{U}_A$ is named the quasipotential (see Appendix \ref{app:path_large_deviation}).

In principle, the quasipotential can be computed from the dynamics, starting from the large deviation Hamiltonian, see for instance Appendix \ref{sec:app-quasipotential}. Formula \eqref{eq:quasipotential_variational} gives an expression for the quasipotential, in the cases when the wave kinetic equation has a single attractor. However, when one knows explicitly the microscopic stationary distribution, for instance in the case of equilibrium statistical mechanics, one can compute directly the quasipotential. It is then  related to the entropy. Those different expressions have to give consistent results. Indeed, one may see a
simple example of the relation between microcanonical entropy and the quasipotential for the dilute gas dynamics in \cite{bouchet2020boltzmann}. The case of spatially homogeneous weakly nonlinear wave dynamics has been discussed in \cite{GBE}. For the present problem, we compute the quasipotential from a microcanonical ensemble for the original Schr\"odinger equation model in Appendix \ref{sec:MCE}. One obtains
\begin{align}\label{eq:quasipotential_expression}
\mathcal{U}_A[n] = \begin{cases}
- \int d\boldsymbol{x} d\boldsymbol{p} \log \left( \dfrac{n(\boldsymbol{x}, \boldsymbol{p})}{n_h^{A}(\boldsymbol{p})} \right) & \mbox{if} \quad \mathcal{A}_\omega[n] = A(\omega)  \\
+ \infty & \mbox{otherwise}
\end{cases}.
\end{align}
This result is technically not obvious. As far as we know, it had never been derived before.

As shown in Appendix \ref{sec:quasipotential_lyapunov}, it is a generic property of the quasipotential to play the role of a Lyapunov function for the deterministic relaxation dynamics, in this case the wave kinetic equation.  Such a property can be derived generically from the existence of a large deviation principle, independently on the specific form of the large deviation Hamiltonian. One can note that the quasipotential \eqref{eq:quasipotential_expression} is the opposite of the entropy $S$ (\ref{eq:quasipotential-unnorm}), up to an additive constant. The quasipotential now depends on $A(\omega)$ and satisfies the normalization condition, $\min_{n} \mathcal{U}_A[n] = 0$, where the minimum is achieved when $n(\boldsymbol{x}, \boldsymbol{p}) = n_h^{A} (\boldsymbol{p})$ \eqref{eq:stationary_WKE}.

From Eq. \eqref{eq:scattering_Hamiltonian_DB} and the fact that $\mathcal{U}_A$ is equal to $-S$ up to a constant, it is immediately checked that $\mathcal{U}_A$ solves the stationary Hamiltonian-Jacobi equation,
 \begin{align} \label{eq:Hamiltonian_Jacobi_main}
\mathcal{H} \left[ n, \frac{\delta \mathcal{U}_A}{\delta n} \right] = 0 .
\end{align}

\subsubsection{Time-reversal symmetry and detailed balance}\label{sec:detailed_balance}

Finally, we consider the time-reversal symmetry of the dynamics (see Appendix \ref{app:detailed_balance}). For the present wave kinetic theory, we showed at the end of section \ref{subsec:kinetic_equation} that the time-reversed local spectral density is defined as $n_R(\bx, \bp, t)=n(\bx, -\bp, -t)$, namely that wave vector needs to change sign in addition to time-reversal $t \to -t$. Therefore, it is useful to introduce the involution operator $I$ such that $I[n(\boldsymbol{x}, \boldsymbol{p})] = n(\boldsymbol{x}, - \boldsymbol{p})$.

Since the wave kinetic dynamics is an equilibrium dynamics whose stationary state is characterised by the microcanonical quasipotential (\ref{eq:quasipotential_expression}) at the large deviation level, we expect the fluctuating dynamics to be time-reversible. Following Appendix \ref{app:detailed_balance}, we indeed check that the Hamiltonian satisfies the detailed-balance condition
\begin{align} \label{eq:detailed_balance}
\mathcal{H} \left[ n, \lambda + \frac{\delta \mathcal{U}_A}{\delta n} \right] = \mathcal{H} \left[ I[n], - I[\lambda] \right] ,
\end{align}
thus proving the time-reversal symmetry of the dynamics.
For a trajectory that follows the wave kinetic equation, $\mathcal{U}_A$ monotonically decreases towards 0, and therefore the local spectral density irreversibly approaches the homogeneous distribution $n_h^A$. However, when $\mu$ is small but finite, there remains a possibility that $\mathcal{U}_A$ increases, namely that the spectrum moves away from the attractor $n_h^{A}$. This property is quantified by the fluctuation relation that is equivalent to the detailed balance relation (\ref{eq:detailed_balance})
\begin{align} \label{eq:fluctuation_relation}
\frac{\mathbb{P}_{n(0)}[\left\{ n^\mu(t) = n(t) \right\}_{0 \leqslant t \leqslant t_f}]}{\mathbb{P}_{n(t_f)}[\left\{ n^\mu (t) = n_R(t - t_f)\right\}_{0 \leqslant t \leqslant t_f}]} \underset{\mu \to 0}{\asymp} \exp \left(  \frac{\mathcal{U}_A[n(0)] - \mathcal{U}_{A}[n(t_f)]}{(2 \pi \mu)^d} \right) .
\end{align}

The fluctuation relation (\ref{eq:fluctuation_relation}) is the fundamental answer to the irreversibility paradox that arises in any classical kinetic theory, and in particular for the classical wave kinetic theory. Indeed, let us consider a path starting from $n(0)$ that follows the wave kinetic equation until it reaches a state $n(t_f)$ at some time $t_f>0$. Since the quasipotential $\mathcal{U}_A$ is a Lyapunov function, one has $\mathcal{U}_A[n(t_f)]<\mathcal{U}_A[n(0)]$. From the fluctuation relation Eq. (\ref{eq:fluctuation_relation}), the exact time-reversed path starting from the state $n(t_f)$ has a probablity $\sim \exp\left(-(2\pi\mu)^{-d}\Delta \mathcal{U} \right)$ (with $\Delta \mathcal{U} = \mathcal{U}_A[n(0)] - \mathcal{U}_{A}[n(t_f)]>0$) to occur in the small $\mu$ limit. The irreversibility turns into an improbability.

\subsection{Decomposition of the Hamiltonian}
Since a free wave packet does not change its wave vector during a free propagation, and since also wave frequency is conserved during scattering by a time-independent potential, waves with different pulsation $\omega(\bp)=\lvert \bp \rvert^{2} / 2$ (\emph{i.e.} located on distinct spherical shells) do not interfere with each other. Therefore, the dynamics can be separated into an infinite number of subsystems in which the degrees of freedom are mutually independent. Indeed, this expectation can be verified by showing that the Hamiltonian is decomposed as an integration over frequency, as we do now.

To do so, let us rewrite the wave vector as $\boldsymbol{p} = p \boldsymbol{e}_\theta$ and define the corresponding frequency, $\omega = p^2 / 2$. The vector $\boldsymbol{e}_\theta$ is a unit vector whose angle is specified by $\theta$. Generally, the number of degrees of freedom for the angle $\theta$ is $d-1$. For example, for a $d = 3$ case, elevation and azimuthal angles would be selected as a set of representative coordinate variables. The following consideration is not dependent on the choice of these coordinates. We just need to assume that a pair of opposite angles, $\theta$ and $-\theta$, are defined such that $\boldsymbol{e}_\theta = - \boldsymbol{e}_{-\theta}$. We decompose the wave vector integration element as $d\boldsymbol{p} = p^{d-1} dp d\theta = (2 \omega)^{(d-2)/2} d\omega d\theta$ with $d\theta$ a surface element on a $(d-1)$-dimensional unit sphere, $S^{d-1}$. We define new variables labeled by frequency $\omega$ as
\begin{subequations}
\begin{align}
n^\omega(\boldsymbol{x}, \theta) & = p^{d-2} n(\boldsymbol{x}, p \boldsymbol{e}_\theta) \\
\lambda^\omega (\boldsymbol{x}, \theta) & = \lambda(\boldsymbol{x}, p \boldsymbol{e}_\theta) \\
\sigma^\omega (\theta_1, \theta_2) & = 2 \pi \Pi (p(\boldsymbol{e}_{\theta_1} - \boldsymbol{e}_{\theta_2}))
\end{align}
\end{subequations}
Please do not confuse $n^\omega$ with $n^\mu$, in the context of this section. The Hamiltonian is decomposed into independent subdynamics, each of which involves the new variables labeled by $\omega$,
\begin{subequations}
\begin{align}
\mathcal{H} [n, \lambda] & = \int \mathcal{H}^\omega [n^\omega, \lambda^\omega] d\omega = \int (\mathcal{H}^\omega_F + \mathcal{H}^\omega_S) d\omega \\
\mathcal{H}^\omega_F [n^\omega, \lambda^\omega] & = - \int d\boldsymbol{x} d\theta (2 \omega)^{1/2} \lambda^\omega (\boldsymbol{x}, \theta) \boldsymbol{e}_\theta \cdot \nabla_x n^\omega (\boldsymbol{x}, \theta) \\
\mathcal{H}^\omega_S [n^\omega, \lambda^\omega] & = c \int d\boldsymbol{x} d\theta_{12} \nonumber \\
& \times \Biggl[ (2 \omega)^{(d-2)/2} (\lambda^\omega (\boldsymbol{x}, \theta_1) - \lambda^\omega (\boldsymbol{x}, \theta_2)) \sigma^\omega (\theta_1, \theta_2) n^\omega(\boldsymbol{x}, \theta_2) \nonumber \\
& + \frac{1}{2} (\lambda^\omega (\boldsymbol{x}, \theta_1) - \lambda^\omega (\boldsymbol{x}, \theta_2))^2 \sigma^\omega (\theta_1, \theta_2) n^\omega(\boldsymbol{x}, \theta_1) n^\omega(\boldsymbol{x}, \theta_2) \Biggr] ,
\end{align}
\end{subequations}
where the integration for $(\boldsymbol{x}, \theta)$ is carried out over $\Gamma \times S^{d-1}$. For this system, the diffusion kernel and the Lyapunov function are
\begin{subequations}
\begin{align}
\Sigma^\omega[n^\omega](\bx; \theta_1, \theta_2) & = -c \sigma^\omega(\theta_1, \theta_2) n^\omega(\bx, \theta_1) n^\omega(\bx, \theta_2) \nonumber \\
& + c \int d\theta' \sigma^\omega(\theta_1, \theta') n^\omega(\bx, \theta_1) n^\omega(\bx, \theta') \delta(\theta_1 - \theta_2) \\
{\rm and} \quad S^\omega[n^\omega] & = \int d\boldsymbol{x} d\theta (2 \omega)^{(d-2)/2} \log n^\omega(\boldsymbol{x}, \theta) .
\end{align}
\end{subequations}
The scattering part of the Hamiltonian is accordingly represented as
\begin{align}
\mathcal{H}^\omega_S & = \int d\bx d\theta_{12} \lambda^\omega (\bx, \theta_1) \Sigma[n^\omega](\bx; \theta_1, \theta_2) \left( \frac{\delta S^\omega}{\delta n^\omega (\bx, \theta_2)} + \lambda^\omega (\bx, \theta_2) \right) .
\end{align}

We may compute the probability of a path of local spectral density separately for each frequency band. For each subsystem, the  amount of wave action, $\propto \int d\boldsymbol{x} d\theta n^\omega (\boldsymbol{x}, \theta) \equiv \mathcal{N} [n^\omega]$, is invariant. For the attractor of the wave kinetic equation, namely the homogeneous distribution, $n^\omega$ is everywhere constant in $\Gamma \times S^{d-1}$, which we write $n^\omega_h$. Finally, we find that the quasipotential,
\begin{align}
\mathcal{U}^\omega [n^\omega] = \begin{cases}
- \int d\boldsymbol{x} d\theta (2 \omega)^{(d-2)/2} \log \left( \dfrac{n^\omega(\boldsymbol{x}, \theta)}{n^\omega_h} \right) & \mbox{if} \quad \mathcal{N} [n^\omega] = \mathcal{N} [n^\omega_h]  \\
+ \infty & \mbox{otherwise}
\end{cases}
\end{align}
satisfies the detailed balance condition, $\mathcal{H}^\omega[n^\omega, \lambda^\omega + \delta \mathcal{U}^\omega / \delta n^\omega] = \mathcal{H}^\omega[\mathcal{I}[n^\omega], - \mathcal{I}[\lambda^\omega]]$ with the involution $\mathcal{I}$ defined as $\mathcal{I}[n^\omega(\boldsymbol{x}, \theta)] = n^\omega(\boldsymbol{x}, -\theta)$. Consequently, the fluctuation relation for a path and its reverse applies to each subsystem separately.

\subsection{Diffusive limit}\label{sec:diffusive_limit}

In this paper, we have focused on the wave kinetic regime when the random potential presents high oscillations at a scale comparable to a typical wavelength of the wave field. Another relevant limit, referred to as the diffusive approximation \cite{kafiabad2019diffusion} or the Fokker-Planck limit \cite{bal2010kinetic}, corresponds to the regime when the spatial oscillations of the potential are at a scale larger than those of the wave field. In this regime, a wave signal propagates along rays which are randomly \emph{refracted} by the potential, leading to an asymptotic diffusion equation for the local spectral density \cite{kafiabad2019diffusion, bal2010kinetic}.

Technically, the diffusion equation on the local spectral density is often derived from a multiscale expansion \yo{of} the microscopic dynamics (\ref{eq:schrodinger_scaled}) \cite{kafiabad2019diffusion, bal2010kinetic}. But interestingly, the diffusive limit can also be obtained from the scattering kinetic regime that we have considered in this paper. Our goal in this subsection is to show how one can derive a path large deviation theory for wave kinetics in the diffusive regime from the large deviation Hamiltonian \eqref{eq:scattering_Hamiltonian_DB}. In a recent paper, a similar weak scattering limit has been considered to derive the path large deviation principle for plasma below the Debye length, related to the Landau equation, from the path large deviation principle for dilute gases, related to the Boltzmann equation \cite{feliachi2021dynamical}.

In the diffusive regime, the random potential has typical variations over large lengthscales compared to the typical wavelength of the signal. As a consequence, the spectrum of the potential \eqref{eq:def:potential_spectrum} is localised around $\boldsymbol{p}=0$, which implies from the definition of the cross section \eqref{eq:def:cross_section} that incoming wave vectors are randomly refracted by an infinitesimal amount at each time step by the potential. One thus expects to obtain the path large deviation theory of wave transport in the diffusive regime from the scattering regime by assuming that the potential spectrum $\Pi$ is supported in the vicinity of $\bp \approx 0$.

In order to derive the diffusive limit from the large deviation Hamiltonian, we use Eq. \eqref{eq:scattering_Hamiltonian_DB}. We show in Appendix \ref{app:diffusive_limit} that the diffusion kernel $\Sigma$ transforms into a differential operator such that for any test functions $f$ and $g$:
\begin{equation}
  \label{eq:kernel_Sigma_diffusive_limit}
  \int d \bp_1 d \bp_2 f(\bp_1)\Sigma(\bx; \bp_1,\bp_2)g(\bp_2) \approx \int \!\! d \bp \, \nabla_p f(\bp) \cdot \left[ n(\bx, \bp)^{2}\boldsymbol{D}(\bp) \right] \cdot \nabla_p g(\bp) .
\end{equation}
Here, $\boldsymbol{D}$ is the diffusion matrix computed from the potential spectrum $\Pi$ as
\begin{equation}
  \label{eq:def:diffusion_matrix}
  \boldsymbol{D}(\bp) = -\frac{c}{2}\int_{\mathbb{R}} \nabla_{y} \otimes \nabla_{y} R(\bp s) ds \quad \text{and} \quad
  R(\boldsymbol{y}) = \int \!\! d \boldsymbol{\eta} \, \Pi(\boldsymbol{\eta}) \mathrm{e}^{i\boldsymbol{\eta}\cdot\by} .
\end{equation}
From its definition, $R(\boldsymbol{y})$ corresponds to the rescaled two-point correlation function of the random potential.

Using Eq. (\ref{eq:kernel_Sigma_diffusive_limit}), the large deviation Hamiltonian in the diffusive limit reads as
\begin{align}
  \label{eq:large_dev_H_diffusive_limit}
  H[n,\lambda] & =  \int \!\! d\bx d\bp \, \lambda(\bx, \bp) \bp \cdot \nabla_x n(\bx, \bp) \nonumber \\
  & + \int \!\! d\bx d\bp \, \nabla_p \lambda(\bx, \bp) \cdot \left[ n(\bx, \bp)^{2}\boldsymbol{D}(\bp) \right] \cdot \nabla_p \left( \frac{\delta S}{\delta n(\bx, \bp)} + \lambda(\bx, \bp) \right) .
\end{align}
The diffusive Hamiltonian (\ref{eq:large_dev_H_diffusive_limit}) conserves the spectral density at a given pulsation $\omega(\bp) = \lvert \bp \rvert^{2} / 2$ because of the fundamental property
\begin{equation}
  \label{eq:orthogonality_D-p}
  \boldsymbol{D}(\bp)\cdot \bp = 0 ,
\end{equation}
which results from $\bp \cdot \nabla R(\bp s) = \partial R(\bp s) / \partial s$. This property means that the diffusion of the spectrum is always orthogonal to the group velocity.

Moreover, the expression of the Hamiltonian (\ref{eq:large_dev_H_diffusive_limit}) makes it clear that the quasipotential of the diffusive dynamics is the same as the quasipotential of the scattering limit Eq. (\ref{eq:quasipotential_expression}). Furthermore, one can easily check that the detailed balance relation \eqref{eq:detailed_balance} is still satisfied by the Hamiltonian (\ref{eq:large_dev_H_diffusive_limit}).

\section{Conclusions and perspective} \label{sec:perspective}
The linear wave kinetic equation is a statistical model that governs wave action density spreading in position and wave-vector space through propagation and scattering in random media. Motivated by recent works on dynamical large deviation principles for kinetic theory, this study has derived the large deviation principle describing the probability of a finite-time evolution of the local spectral density of wave action in an asymptotic limit of scale separation. Importantly, the large deviation principle that is derived in this paper satisfies a time-reversal symmetry with respect to the microcanoical quasipotential, that is directly (and independently) computed from the microcanonical measure.

In this paper, we restrict our considerations to the simplest Schr\"odinger model with a homogeneous random potential. The next step is to extend the present formulation to a wider range of situations. Possible technical difficulties encountered during such future works involve (i) generalization of the dispersion relation to a function of position and wave vector, $\omega(\boldsymbol{x}, \boldsymbol{p})$, (ii) coping with spatial inhomogeneity for the randomness, that leads to a space-dependent scattering cross section, $\sigma(\boldsymbol{x}, \boldsymbol{p}_1, \boldsymbol{p}_2)$, (iii) consideration of a vector field that involves polarized waves or multiple waves, e.g., elastic media holding compressional and shear waves. Combining the present approach and a previous work on wave turbulence \cite{GBE}, we have also derived a formula of path-large deviations for inhomogeneous spectral density in nonlinear 4-wave interacting systems \cite{GOF}. \yo{Interestingly, combination of weak non-linearity and random potential can lead to a new kinetic regime in dimension one \cite{nazarenko2019wave}. The generalisation of the large deviation theory to this case remains to our knowledge an open question and is left for further work.}

In the context of geophysics, the wave kinetic equation is used to discuss energy cascades of internal waves in the oceans and atmosphere where rotation and density stratification play key roles \cite{savva2018scattering}. Because the dispersion relation of internal waves depends not on \yo{the magnitude of a} wave vector but on its angle against the gravity direction, even linear theory predicts interscale energy transfer. In this process, balanced geostrophic turbulent flow acts as a random potential. If we assume a stationary flow state, i.e., fix the potential field in time, the formulation will be analogous to the present case. On the other hand, once we allow the temporal variations in the geostrophic flow field, the situation essentially changes. Wave frequency is no longer conserved during propagation and scattering. Spreading of action density in frequency space associates gain or loss of wave energy. Quantification of the energy exchange rate between evolving turbulent eddies and waves remains an open problem. Recently, Dong et al. \cite{dong2020frequency} discussed wave frequency diffusion in geostrophic turbulence based on a kind of kinetic equation model. As pointed out in the present study, the ordinary kinetic equation predicts an irreversible change in the spectral density. In the actual environmental situation, the scale separation parameter, $\mu$, is not necessarily small and there should be non-negligible fluctuations in spectral density. The large deviation formulation has a possible application for the estimation of a fluctuating energy transfer rate, in such regime where the kinetic theory is marginally valid.

\bmhead{Acknowledgments}
This work was supported by JSPS Overseas Research Fellowship as well as KAKENHI Grant Number JP20K14556 (Y.O.), and by the Simons Foundation through the Collaboration Grant 651463 “Wave Turbulence” (F.B. and J.G.) and the Targeted Grant in MPS 663054 “Revisiting the Turbulence Problem Using Statistical Mechanics” (F.B.). We thank Gregory Eyink, Laure Saint-Raymond, Jacques Vanneste, and Antoine Venaille for fruitful discussions.

\appendix
\section{Properties of the stochastic system specified by a large deviation Hamiltonian} \label{sec:fluctuation}

\subsection{Path large deviation}\label{app:path_large_deviation}
This appendix presents some general properties of a stochastic process $X^\epsilon(t)$ whose probability conditioned on an initial value, $X^\epsilon(t_i) = X(t_i)$, is specified at the large deviation level via a formula,
\begin{subequations} \label{eq:PLD}
\begin{gather}
\mathbb{P} \left[ \left\{ X^\epsilon(t) = X(t) \right\}_{t_i \leqslant t \leqslant t_f} \right] \underset{\epsilon \to 0}{\asymp} \exp \left( - \frac{\mathcal{S}[X]}{\epsilon}  \right) \\
\mathcal{S}[X] = \int_{t_i}^{t_f} dt \mathcal{L}(X, \dot{X}) \equiv \int_{t_i}^{t_f} dt \sup_P \left\{ P \cdot \dot{X} - \mathcal{H}(X, P) \right\} .
\end{gather}
\end{subequations}
Here, $X^\epsilon(t)$ can be a scalar, vector, or continuous function defined on some space. Basic requirements are that an inner product is properly defined, and the dynamical property of the system is controlled by a single non-negative parameter $\epsilon$. For the simplicity, we regard $X$ as a scalar but the following consideration can be immediately extended to general cases. The large deviation Hamiltonian $\mathcal{H}(X, P)$ is a convex function of $P$ and satisfies $\mathcal{H}(X, 0) = 0$ for any $X$. From the definition, the Lagrangian $\mathcal{L}$ satisfies $\mathcal{L}(X, \dot{X}) \geq P \cdot \dot{X} - \mathcal{H}(X, P)$ for any $X$, $\dot{X}$ and $P$. Therefore, inserting $P=0$, we know $\mathcal{L} \geq 0$.

\subsubsection{Relaxation path}
\label{app:sub:relaxation_path}
Clearly, in the limit of $\epsilon \to 0$, the system becomes deterministic with a single path that minimizes the action $\mathcal{S}[X]$ for a prescribed initial condition $X(t_i)$\textemdash named the relaxation path. Since $\mathcal{L} \geq 0$, if there exists a function $R(X)$ that satisfies $\mathcal{L}(X, R(X)) = 0$, a path solving $\dot{X} = R(X)$ minimizes the action and yields $\min_{X} \mathcal{S}[X] = 0$. From the facts that $\mathcal{L}(X, \dot{X}) = P \cdot \dot{X} - \mathcal{H}(X, P)$ with $P$ solving $\dot{X} - \partial \mathcal{H} / \partial P = 0$ and $\mathcal{H}(X, 0) = 0$ for any $X$, we understand that a function $R(X) = \left. \partial \mathcal{H} / \partial P \right\vert_{P=0}$ fulfills $\mathcal{L}(X, R(X)) = 0$. We thus assert that an equation
\begin{align} \label{eq:relaxation}
\dot{X}^r = R(X^r) \equiv \frac{\partial \mathcal{H}}{\partial P} (X^r, 0)
\end{align}
determines the relaxation path $X^r(t)$.

\subsubsection{Optimal path}
Slightly changing the situation, if we fix both the initial and final states, $X(t_i) = x_i$ and $X(t_f) = x_f$, respectively, the most probable path from $x_i$ to $x_f$, namely the optimal path, or the instanton, is obtained by again minimizing $\mathcal{S}[X]$. This problem is equivalent to the principle of least action in analytical mechanics. In this context, $P$ is called the generalized \yo{momentum} and represented by $P = \partial \mathcal{L} / \partial \dot{X}$ which is no longer 0. The optimal path in phase space is governed by a set of canonical equations,
\begin{subequations} \label{eq:canonical}
\begin{align}
\dot{X} & = \frac{\partial \mathcal{H}}{\partial P} \\
\dot{P} & = - \frac{\partial \mathcal{H}}{\partial X} ,
\end{align}
\end{subequations}
and we shall write their solutions as $X^o[x_f, t_f; x_i, t_i]$ and $P^o[x_f, t_f; x_i, t_i]$. For the simplicity, we fix the initial conditions, $t_i$ and $x_i$, and rewrite the final state as $x$ and $t$. We then introduce the Hamilton's principal function $\mathcal{Q}$ as an integration of the action following the optimal path as
\begin{align}
  \mathcal{Q}(x, t) = \int_{t_i}^t d\tau \mathcal{L} (X^o(\tau), \dot{X}^o(\tau)) .
\end{align}
It is known in analytical mechanics that in this case the generalized momentum at the final time is represented as $P^o(t) = \left. \partial \mathcal{L} / \partial \dot{X} \right\vert_{t} = \partial \mathcal{Q} / \partial x$, and $\mathcal{Q}$ solves the Hamilton-Jacobi equation,
\begin{align} \label{eq:Hamilton_Jacobi}
\frac{\partial \mathcal{Q}}{\partial t} + \mathcal{H}\left( x, \frac{\partial \mathcal{Q}}{\partial x} \right) = 0 .
\end{align}
In the definition of $\mathcal{Q}(x, t)$, a set of arguments, $(x, t)$, is arbitrary chosen. When we pick up an optimal path $\left\{ X^o(\tau), P^o(\tau) \right\}$, at any point on this path, the generalized momentum $P$ and the Hamilton's principle function $\mathcal{Q}$ is related via
\begin{align} \label{eq:generalized_momenta}
P^o(\tau) = \frac{\partial \mathcal{Q}}{\partial x} (X^o(\tau), \tau) .
\end{align}
Therefore, combining (\ref{eq:generalized_momenta}) and the first line of (\ref{eq:canonical}), we formally obtain a single equation determining the optimal path,
\begin{align} \label{eq:optimal_path}
\frac{d X^o}{d \tau} = \frac{\partial \mathcal{H}}{\partial P} \left( X^o, \frac{\partial \mathcal{Q}}{\partial x} (X^o, \tau) \right) .
\end{align}
This equation is, however, not generally useful because $\mathcal{Q}(x, t)$ is inaccessible in most cases.

\subsubsection{Quasipotential and fluctuation path} \label{sec:app-quasipotential}
Going back to the original stochastic model, the meaning of \yo{$\mathcal{S}$} is understood as the rate function for the probability that $X^\epsilon$ reaches $x_f$ at $t = t_f$. Indeed, we may derive from (\ref{eq:PLD}) an expression
\begin{align} \label{eq:rate_function_S}
\mathbb{P} \left[ X^\epsilon(t) = x \vert X^\epsilon(t_i) = x_i \right] \underset{\epsilon \to 0}{\asymp} \exp \left( - \frac{\mathcal{Q}(x, t)}{\epsilon}  \right)
\end{align}
based on the contraction principle. Now, we shall consider the stationary distribution of the probability density of $X^\epsilon$. This can be done simply setting $t_i = -\infty$ in (\ref{eq:rate_function_S}). To make the discussion more specific, let us assume that the relaxation dynamics (\ref{eq:relaxation}) has a unique global attractor $x_0$, where $R(x_0) = 0$. Then, we set $x_i = x_0$ and also $t = 0$, to write a large deviation formula for the stationary distribution
\begin{align}
\mathbb{P}_S (x) \underset{\epsilon \to 0}{\asymp} \exp \left( - \frac{\mathcal{U}(x)}{\epsilon}  \right)
\end{align}
with
\begin{align} \label{eq:quasipotential_variational}
\mathcal{U}(x) = \inf_{X(t) \vert X(-\infty) = x_0 \mbox{ and } X(0) = x} \int_{-\infty}^0 dt \mathcal{L} (X(t), \dot{X}(t)) .
\end{align}
The rate function $\mathcal{U}$ is called the quasipotential. Since $\mathcal{U}$ is the special case of $\mathcal{Q}$ but independent of $t$, it solves the stationary version of the Hamilton-Jacobi equation (\ref{eq:Hamilton_Jacobi}),
\begin{align} \label{eq:stationary_Hamilton_Jacobi}
\mathcal{H} \left( x, \frac{\partial \mathcal{U}}{\partial x} \right) = 0 .
\end{align}
For the present case, the optical path $X^o(\tau)$ represents the most probable route from an attractor $x_0$ to a specific point $x$. This route is called the fluctuation path and is denoted by $X^f(t)$. Once we obtain the quasipotential $\mathcal{U}(x)$, (\ref{eq:optimal_path}) provides an equation determining the fluctuation path as
\begin{align}
\dot{X}^f = F(X^f) \equiv \frac{\partial \mathcal{H}}{\partial P} \left( X^f, \frac{\partial \mathcal{U}}{\partial x} (X^f) \right) .
\end{align}
Since the vector field $F(x)$ does not depend on $t$, this equation is more useful than the original one (\ref{eq:optimal_path}). On the fluctuation path, the generalized momentum is computed based on (\ref{eq:generalized_momenta}) as $P^f = \partial \mathcal{U} / \partial x (X^f)$. Combining (\ref{eq:stationary_Hamilton_Jacobi}) with the fact that $\mathcal{H}$ is constant along the optical path, we understand that $\mathcal{H}(X^f, P^f) = 0$ always holds.

\subsubsection{Quasipotential as a Lyapunov function}\label{sec:quasipotential_lyapunov}
A relaxation path and a fluctuation path have distinct properties for the variations in $\mathcal{U}$. For a relaxation path, we have
\begin{align*}
\frac{d \mathcal{U}}{d t}(X^r) & = \dot{X}^r \frac{\partial \mathcal{U}}{\partial x}(X^r) = \frac{\partial \mathcal{H}}{\partial P} (X^r, 0) \frac{\partial \mathcal{U}}{\partial x}(X^r) \\
& = \underbrace{\mathcal{H}(X^r, 0)}_{= 0} - \underbrace{\mathcal{H} \left( X^r, \frac{\partial \mathcal{U}}{\partial x} (X^r) \right)}_{= 0} + \frac{\partial \mathcal{H}}{\partial P} (X^r, 0) \frac{\partial \mathcal{U}}{\partial x}(X^r) \leq 0 ,
\end{align*}
where we have used the general expressions, $\mathcal{H}(X, 0) = \mathcal{H}(X, \partial \mathcal{U} / \partial x (X)) = 0$, and the convexity of $\mathcal{H}(X, P)$ for $P$. For a fluctuation path, we have
\begin{align*}
\frac{d \mathcal{U}}{d t}(X^f) & = \dot{X}^f \frac{\partial \mathcal{U}}{\partial x}(X^f) = \frac{\partial \mathcal{H}}{\partial P} \left( X^f, \frac{\partial \mathcal{U}}{\partial x}(X^f) \right) \frac{\partial \mathcal{U}}{\partial x}(X^f) \\
& = \underbrace{\mathcal{H}(X^f, 0)}_{= 0} - \underbrace{\mathcal{H} \left( X^f, \frac{\partial \mathcal{U}}{\partial x} (X^f) \right)}_{= 0} + \frac{\partial \mathcal{H}}{\partial P} \left( X^f, \frac{\partial \mathcal{U}}{\partial x}(X^f) \right) \frac{\partial \mathcal{U}}{\partial x}(X^f) \geq 0 ,
\end{align*}
again from the convexity of $\mathcal{H}$. We have thus learned that the quasipotential is a Lyapunov function because it monotonically decreases in a relaxation path while increases in a fluctuation path. These results are natural consequence from a basic property that $\mathcal{U}(x)$ is minimum at the attractor $x_0$ and the relaxation and fluctuation paths represent routes to and from the attractor.

\subsection{Properties of large deviation Hamiltonian}
\subsubsection{Conservation law} \label{sec:conservation_appendix}
From now on, we shall regard $X$ and $P$ as vectors so that there are multiple directions in $X$ space. For this case, when the Hamiltonian $\mathcal{H}$ possesses a kind of symmetry, it is related to the conservation law of the system. More specifically, let us suppose that we find a function $\mathcal{C}(X)$ that satisfies
\begin{align} \label{eq:Hamiltonian_conservation}
\mathcal{H} \left( X, P + \alpha \frac{\partial \mathcal{C}}{\partial X} \right) = \mathcal{H} (X, P)
\end{align}
for any $X$, $P$, and $\alpha$. This condition is equivalent to
\begin{align*}
\frac{\partial \mathcal{H}}{\partial P}(X, P) \cdot \frac{\partial \mathcal{C}}{\partial X} (X) = 0
\end{align*}
for any $X$ and $P$. Now that $\mathcal{H}(X, \cdot)$ is flat in the direction of $\partial \mathcal{C} / \partial X$, from the property of the Legendre-Fenchel transform \cite{touchette2009large}, the corresponding Lagrangian has a property,
\begin{align}
\mathcal{L} (X, \dot{X}) = + \infty \quad \mbox{if} \quad \dot{X} \cdot \frac{\partial \mathcal{C}}{\partial X} (X) \neq 0 .
\end{align}
This expression indicates that the probability for a path crossing a contour of $\mathcal{C}$ is strictly 0. This constraint applies not only to the optimal path but also to any path with random fluctuations. We thus understand that (\ref{eq:Hamiltonian_conservation}) serves as a condition for $\mathcal{C}$ being an invariant of the system.

\subsubsection{Detailed balance}\label{app:detailed_balance}
The detailed balance is a property of equilibrium states which asserts time-reversibility of the process, meaning that the probabilities of any trajectory and its reversed counterpart are equal. A basic expression of detailed balance for a stationary stochastic process is $\mathbb{P}_{\Delta t}(y; x) \mathbb{P}_S(x) = \mathbb{P}_{\Delta t}(x; y) \mathbb{P}_S(y)$, where $\mathbb{P}_{\Delta t}(y; x)$ is the transition probability from a state $x$ to another state $y$ during a time interval $\Delta t$, and $\mathbb{P}_S(x)$ is the stationary probability distribution.

Since we are now considering a continuous Markov process, it is enough to regard $\Delta t$ as arbitrary small. For the limit of $\Delta t \to 0$, we may write $y \sim x + \dot{x} \Delta t$ and redefine the transition probability as $\mathbb{P}_{\Delta t}(x, \dot{x}) \sim \mathbb{P}_{\Delta t}(x + \dot{x} \Delta t; x)$. Assuming the continuity of $\mathbb{P}$ and $\mathbb{P}_S$, the detailed balance condition is rewritten as
\begin{align} \label{eq:detailed_balance_1}
\mathbb{P}_{\Delta t}(x, \dot{x}) \mathbb{P}_S(x) \sim \mathbb{P}_{\Delta t}(x + \dot{x} \Delta t, - \dot{x}) \mathbb{P}_S(x + \dot{x} \Delta t) .
\end{align}
For the present problem, the probability distribution is specified as $\mathbb{P}_{\Delta t}(x, \dot{x}) \asymp \exp(- \Delta t \mathcal{L}(x, \dot{x}) / \epsilon)$ and $\mathbb{P}_S (x) \asymp \exp(- \mathcal{U}(x) / \epsilon)$. Therefore, the detailed balance condition (\ref{eq:detailed_balance_1}) is rewritten as
\begin{align}
\mathcal{L}(x, \dot{x}) - \mathcal{L}(x, - \dot{x}) = \dot{x} \cdot \frac{\partial \mathcal{U}}{\partial x} .
\end{align}
This condition is modified in terms of the Hamiltonian via the Legendre-Fenchel transform as
\begin{align} \label{eq:detailed_balance_2}
\mathcal{H}(x, -p) = \mathcal{H} \left( x, p + \frac{\partial \mathcal{U}}{\partial x} \right) .
\end{align}
Because $\mathcal{H}(x, 0) = 0$ in the current problem, the stationary Hamilton-Jacobi equation (\ref{eq:stationary_Hamilton_Jacobi}) is a necessary (but not sufficient) condition for $\mathcal{U}$ being the quasipotential.

Once the detailed balance condition (\ref{eq:detailed_balance_2}) is verified, we understand that the probabilities of a path and its reverse are related at the large deviation level via the expression,
\begin{align}
\frac{\mathbb{P} \left[\left\{ X^\epsilon(t) = X(t) \right\}_{t_i \leqslant t \leqslant t_f} \right]}{\mathbb{P}\left[\left\{ X^\epsilon(t) = X(t_f + t_i - t) \right\}_{t_i \leqslant t \leqslant t_f}\right]} \underset{\epsilon \to 0}{\asymp} \exp \left( - \frac{\mathcal{U}(x_f) - \mathcal{U}(x_i)}{\epsilon} \right) ,
\end{align}
an equivalent form of the Crooks fluctuation theorem. Another outcome of the detailed balance is that the fluctuation path is the time reverse of the relaxation path. This property is derived from
\begin{align*}
R(X) = \frac{\partial \mathcal{H}}{\partial P}(X, 0) = - \frac{\partial \mathcal{H}}{\partial P} \left( X, \frac{\partial \mathcal{U}}{\partial X} \right) = - F(X) .
\end{align*}

\section{Microcanonical ensemble and quasi-potential for the Schr\"odinger equation} \label{sec:MCE}

In this appendix, we consider the microcanonical ensemble of the dynamics governed by the Schr\"odinger equation, (\ref{eq:schrodinger_scaled}). The aim is to compute the quasipotential of the local empirical spectral density, i.e., the large deviation rate function of $n^\mu$, in the small $\mu$ limit. We will prove that
\begin{align}
\label{eq:stationary-2}
\mathbb{P}^\mu_{A,m}[n^\mu = n]  \underset{\mu\to0}{\asymp} \exp \left(-\frac{\mathcal{U}_A[n]}{(2 \pi \mu)^d} \right),
\end{align}
with $\mathbb{P}^\mu_{A,m}$ the probabilities with respect to the microcanonical measure with the constraints,
\begin{align}
\mathcal{A}_\omega [n^\mu] & \equiv \int h \left( \omega - \frac{\vert\bp\vert^2}{2} \right) n^\mu(\boldsymbol{x}, \boldsymbol{p}) d\boldsymbol{x} d\boldsymbol{p} = A(\omega),
\end{align}
where $h$ is the Heaviside function, and $A: \mathbb{R}^+ \to \mathbb{R}^+$ is a prescribed function. The expected form of the rate function, namely, the quasipotential, is
\begin{align}
\label{eq:quasipotential-2}
\mathcal{U}_A[n] = \begin{cases}
- \int d\boldsymbol{x} d\boldsymbol{p} \log \left( \dfrac{n(\boldsymbol{x}, \boldsymbol{p})}{\yo{n^A_h}(\boldsymbol{p})} \right) & \mbox{if} \quad \mathcal{A}_\omega[n] = A(\omega)  \\
+ \infty & \mbox{otherwise}
\end{cases} ,
\end{align}
where
\begin{align} \label{eq:stationary_WKE_2}
n^A_{h}(\boldsymbol{p}) = \dfrac{A'(\lvert \boldsymbol{p} \rvert^2 / 2)}{\int d\boldsymbol{x} d\boldsymbol{\eta} \delta(\lvert \boldsymbol{p} \rvert^2 / 2 - \lvert \boldsymbol{\eta} \rvert^2 / 2)} .
\end{align}
In the following proof, we first define the microcanonical measure and then derive the rate function (\ref{eq:quasipotential-2}).

If we consider an infinite domain with a finite amount of total wave action, $\int_{\mathbb{R}^d} \lvert \psi^\mu \rvert^2 d\boldsymbol{x} < \infty$, since the wave action density will be diluted to absolutely 0 everywhere, an equilibrium state of the microcanonical ensemble does not make sense. Here, we instead assume a spatial periodicity of the scaler field $\psi^\mu(\boldsymbol{x})$ and concentrate our attention on a $d$-dimensional cubic domain, $[0, 2\pi)^d \equiv \Gamma \subset \mathbb{R}^d$. In this setting, the empirical local spectral density~(\ref{eq:Wigner_distribution_scaled}) consists of delta functions in wave-vector space like
\begin{align}
n^\mu(\boldsymbol{x}, \boldsymbol{p}) = \sum_{\boldsymbol{k} \in (1/2) \mathbb{Z}^d} n^\mu_{\boldsymbol{k}} (\boldsymbol{x}) \delta (\boldsymbol{p} - \mu \boldsymbol{k}) ,
\end{align}
where $n^\mu_k (\boldsymbol{x})$ is a discrete form of Wigner distribution adapted to periodic domains, defined as
\begin{align} \label{eq:discrete_wigner}
n^\mu_{\boldsymbol{k}} (\boldsymbol{x}) & = \frac{1}{(2 \pi)^d} \int_\Gamma d\boldsymbol{y} e^{- 2 i \boldsymbol{k} \cdot \boldsymbol{y}} \psi^\mu \left(\boldsymbol{x} + \boldsymbol{y} \right) \psi^{\mu \dag} \left(\boldsymbol{x} - \boldsymbol{y} \right) .
\end{align}
We also introduce the Fourier coefficients of $\psi^\mu$ as
\begin{subequations}
\begin{align}
\hat{\psi}^\mu_{\boldsymbol{k}} & = \frac{1}{(2\pi)^d} \int_\Gamma e^{- {\rm i} \boldsymbol{k} \cdot \boldsymbol{x}} \psi^\mu(\boldsymbol{x}) d\boldsymbol{x} \\
\psi^\mu(\boldsymbol{x}) & = \sum_{\boldsymbol{k} \in \mathbb{Z}^d} e^{{\rm i} \boldsymbol{k} \cdot \boldsymbol{x}} \hat{\psi}^\mu_{\boldsymbol{k}} .
\end{align}
\end{subequations}
Perceval's theorem allows us to write the wave action density per unit volume in three forms,
\begin{align}
\frac{1}{(2\pi)^d} \int_{\Gamma \times \mathbb{R}^d} n^\mu (\boldsymbol{x}, \boldsymbol{p}) d\boldsymbol{x} d\boldsymbol{p} = \frac{1}{(2\pi)^d} \int_{\Gamma} \lvert \psi^\mu(\boldsymbol{x}) \rvert^2 d\boldsymbol{x} = \sum_{\boldsymbol{k}} \lvert \hat{\psi}^\mu_{\boldsymbol{k}} \rvert^2 .
\end{align}
We shall set $\mu \to 0$ while keeping this action density finite. When we fix a volume element in $\bp$ space, the number of $\bk$ vectors which are involved there increases as $\sim \mu^{-d}$. Therefore, the typical amplitude of the Fourier coefficients depends on $\mu$ as $\hat{\psi}^\mu_{\boldsymbol{k}} \sim \mathcal{O}(\mu^{d/2})$. Now we scale $\hat{\psi}^\mu$ by introducing a new coefficient, $a^\mu_{\boldsymbol{p}}$ with $\boldsymbol{p} \in \mu \mathbb{Z}^d$, as $\hat{\psi}^\mu_{\boldsymbol{k}} = \mu^{d/2} a^\mu_{\mu \boldsymbol{k}}$, \yo{i.e.},
\begin{align}
a^\mu_{\boldsymbol{p}} = \frac{1}{(2 \pi \mu^{1/2})^d} \int_\Gamma e^{- {\rm i} \boldsymbol{p} \cdot \boldsymbol{x} / \mu} \psi^\mu(\boldsymbol{x}) d\boldsymbol{x} .
\end{align}
Note that $a^\mu_{\boldsymbol{p}}$ remains finite for $\mu \to 0$, but its norm, $\sum_{\boldsymbol{p} \in \mu \mathbb{Z}^d} \lvert a^\mu_{\boldsymbol{p}} \rvert^2 = (2\pi \mu)^{-d} \int_{\Gamma \times \mathbb{R}^d} n^\mu (\boldsymbol{x}, \boldsymbol{p}) d\boldsymbol{x} d\boldsymbol{p}$, diverges in the same limit.

To apply equilibrium statistical mechanics, we consider the phase space spanned by the scaled coefficients, $\{ a^\mu_{\boldsymbol{p}} \}_{\boldsymbol{p} \in \mu \mathbb{Z}^d}$. In this space, the Lebesgue measure $\mathfrak{m}$ is represented as
\begin{align}
d \mathfrak{m} = \prod_{\boldsymbol{p}} d a^\mu_{\boldsymbol{p}} d a^{\mu \dag}_{\boldsymbol{p}} \equiv \prod_{\boldsymbol{p}} d a^r_{\boldsymbol{p}} d a^i_{\boldsymbol{p}} ,
\end{align}
where $a^\mu_{\boldsymbol{p}} = (a^r_{\boldsymbol{p}} + {\rm i} a^i_{\boldsymbol{p}}) / \sqrt{2}$ is understood. This measure makes sense only when an upper limit of the wave vector is set to truncate the infinite product. As a result, the number of degrees of freedom of $\psi^\mu$ in physical space is also restricted. Let us define the bounded set of $\boldsymbol{k}$ as $\mathbb{K}_\Delta \equiv \left\{ - 1 / (2\Delta) + 1, \ldots, 1 / (2\Delta) \right\}^d$, and accordingly that of $\boldsymbol{p}$ as $\mu \mathbb{K}_\Delta$. The number of elements in $\mathbb{K}_\Delta$ is $\mathcal{N} \equiv 1 / \Delta^d$. Then, $\psi^\mu$ is specified by the values at $\mathcal{N}$ points, $\Gamma_\Delta \equiv \left\{0, \Delta, \ldots, 2\pi - \Delta \right\}^d$, and the values in $\Gamma \setminus \Gamma_\Delta$ are determined by interpolation. The Lebesgue measure is now represented in either wave vector or position space as
\begin{align}
d\mathfrak{m} = \prod_{\boldsymbol{p} \in \mu \mathbb{K}_\Delta} d a^\mu_{\boldsymbol{p}} d a^{\mu \dag}_{\boldsymbol{p}} = J^\mu_{\Delta} \prod_{\boldsymbol{x} \in \Gamma_\Delta} d \psi^\mu(\boldsymbol{x}) d \psi^{\mu \dag}(\boldsymbol{x}) ,
\end{align}
where $J^\mu_\Delta$ is the Jacobian of the function that maps $a^\mu_{\boldsymbol{p}}$ to \yo{$\psi^\mu$}. To compute this Jacobian, we consider the integral,
\begin{align}
\int d\mathfrak{m} \exp \left[ - \frac{1}{(2\pi\mu)^d} \int_{\Gamma \times \mathbb{R}^d} n^\mu(\boldsymbol{x}, \boldsymbol{p}) d\boldsymbol{x} d\boldsymbol{p} \right]
\end{align}
that we express in two different ways:
\begin{align}
& \int \left( \prod_{\boldsymbol{p} \in \mu \mathbb{K}_\Delta} d a^\mu_{\boldsymbol{p}} d a^{\mu \dag}_{\boldsymbol{p}} \right) \exp \left[ - \sum_{\boldsymbol{p} \in \mu \mathbb{K}_\Delta} \lvert a^\mu_{\boldsymbol{p}} \rvert^2 \right] \nonumber \\
= & J^\mu_\Delta \int \left( \prod_{\boldsymbol{x} \in \Gamma_\Delta} d \psi^\mu(\boldsymbol{x}) d \psi^{\mu \dag}(\boldsymbol{x}) \right) \exp \left[ - \left( \frac{\Delta}{2 \pi \mu} \right)^d \sum_{\boldsymbol{x} \in \Gamma_\Delta} \lvert \psi^\mu (\boldsymbol{x}) \rvert^2 \right] .
\end{align}
The left-hand side turns out to be $(2 \pi)^\mathcal{N}$, and the right-hand side is $J^\mu_\Delta (2 \pi)^\mathcal{N} (2 \pi \mu)^{d\mathcal{N}} \mathcal{N}^\mathcal{N}$. We therefore obtain $J^\mu_\Delta$ and, accordingly,
\begin{align}
d \mathfrak{m} = \prod_{\boldsymbol{x} \in \Gamma_\Delta} \frac{d \psi^\mu(\boldsymbol{x}) d \psi^{\mu \dag}(\boldsymbol{x})}{(2\pi\mu)^d \mathcal{N}} .
\label{Lebesgues}
\end{align}
We denote by $\mathbb{E}$ integrals over the Lebesgue measure (\ref{Lebesgues}). The microcanonical measure with constraints on $A$ is then defined as
\begin{align}
d \mathfrak{m}_A = \frac{\Pi_\omega \delta \left( \mathcal{A}_\omega \left[ n^\mu \right] - A(\omega) \right)}{\mathbb{E}\left[ \Pi_\omega \delta \left( \mathcal{A}_\omega \left[ n^\mu \right] - A(\omega) \right) \right]} d \mathfrak{m},
\end{align}
where $\Pi_\omega \delta \left( \mathcal{A}_\omega \left[ n^\mu \right] - A(\omega) \right)$ means that we constrain the values of all the invariants $\mathcal{A}_\omega$ for any $\omega$.

Our goal is to compute the probability distribution of $n^\mu$ for a microcanonical measure constrained by $A$, i.e.,
\begin{align} \label{eq:microcanonical-distribution}
\mathbb{P}^\mu_{A,m}[n^\mu = n] = \frac{\mathbb{E}\left[ \delta \left( n^\mu - n \right) \Pi_\omega \delta \left( \mathcal{A}_\omega \left[ n^\mu \right] - A(\omega) \right) \right]}{\mathbb{E}\left[ \Pi_\omega \delta \left( \mathcal{A}_\omega \left[ n^\mu \right] - A(\omega) \right) \right]}.
\end{align}
It will be mathematically convenient, for intermediate computations, to use in the following a normalizable Gaussian measure $d \mathfrak{m}_G$,
\begin{align}
d \mathfrak{m}_G & = \prod_{\boldsymbol{p} \in \mu \mathbb{K}_\Delta} e^{- \lvert a^\mu_{\boldsymbol{p}} \rvert^2 } \frac{d a^\mu_{\boldsymbol{p}} d a^{\mu \dag}_{\boldsymbol{p}}}{2 \pi} \nonumber \\
& = \exp \left[ - \frac{1}{(2\pi\mu)^d} \int_{\Gamma \times \mathbb{R}^d} n^\mu (\boldsymbol{x}, \boldsymbol{p}) d\boldsymbol{x} d\boldsymbol{p} - \mathcal{N} \log 2\pi \right] d \mathfrak{m} ,
\end{align}
which satisfies $\int d\mathfrak{m}_G = 1$.  We note that the $\mathcal{N}$-dependent term diverges in the $\Delta \to 0$ limit, but this divergence will be compensated in \yo{$\mathbb{P}_{A, m}^\mu$} below.  We denote $\mathbb{E}_G$ averages with respect to this Gaussian measure.

Then, (\ref{eq:microcanonical-distribution}) can be rewritten as
\begin{align} \label{eq:microcanonical-distribution-2}
& \mathbb{P}^\mu_{A,m}[n^\mu = n] \nonumber \\
= & \dfrac{\exp\left[ (2\pi\mu)^{-d} \int_{\Gamma \times \mathbb{R}^d} n (\boldsymbol{x}, \boldsymbol{p}) d\boldsymbol{x} d\boldsymbol{p} \right] \mathbb{E}_G\left[ \delta \left( n^\mu - n \right) \Pi_\omega \delta \left( \mathcal{A}_\omega \left[ n^\mu \right] - A(\omega) \right) \right]}{\mathbb{E}_G \left\{ \exp\left[ (2\pi\mu)^{-d} \int_{\Gamma \times \mathbb{R}^d} n^\mu (\boldsymbol{x}, \boldsymbol{p}) d\boldsymbol{x} d\boldsymbol{p} \right] \Pi_\omega \delta \left( \mathcal{A}_\omega \left[ n^\mu \right] - A(\omega) \right)  \right\}}.
\end{align}
In the following, when we consider the Gaussian measure $\mathfrak{m}_G$, the continuous limit $\Delta \to 0$ is always understood.

We look for a large deviation principle
\begin{align}
\mathbb{E}_G\left[ \delta \left( n^\mu - n \right) \Pi_\omega \delta \left( \mathcal{A}_\omega \left[ n^\mu \right] - A(\omega) \right) \right] \underset{\mu\to0}{\asymp} \exp \left(-\frac{\mathcal{I}_G[n, A]}{(2 \pi \mu)^d} \right) .
\label{eq:rate_G}
\end{align}
Our strategy is to compute its rescaled cumulant generating function and
to apply the G\"artner-Ellis theorem. We shall then define a free energy as
\begin{align}
f_G[\lambda, \beta] \equiv & - \lim_{\mu \to 0} \left( 2 \pi \mu \right)^d \log \mathbb{E}_G \Biggl\{ \exp \Biggl[ - \frac{1}{\left( 2 \pi \mu \right)^d} \int_{\Gamma \times \mathbb{R}^d} d\boldsymbol{x} d\boldsymbol{p} \nonumber \\
\times & \biggl( \int_{\mathbb{R}^+} \beta(\omega) h \left(\omega - \frac{\lvert \boldsymbol{p} \rvert^2}{2} \right) d\omega + \lambda(\boldsymbol{x}, \boldsymbol{p}) \biggr) n^\mu (\boldsymbol{x}, \boldsymbol{p}) \Biggr] \Biggr\},
\end{align}
where $\beta: \mathbb{R}^+ \to \mathbb{R}$ is a real continuous function representing the chemical potential and $\lambda: \Gamma \times \mathbb{R}^d \to \mathbb{R}$ is also a real continuous function. Because $n^\mu$ is quadratic in $\psi^\mu$, the expectation is a Gaussian integral. To make it explicit, the expression in the square brackets is rewritten as
\begin{align} \label{eq:microcanonical-exponent}
- \frac{1}{\left( 2 \pi \mu \right)^d} \int \psi^{\mu\dag} (\boldsymbol{x}) \mathcal{L}_{\tilde{\lambda}} \psi^\mu (\boldsymbol{x}) d\boldsymbol{x} ,
\end{align}
where $\mathcal{L}_{\tilde{\lambda}}$ is a pseudo-differential operator defined as
\begin{subequations}
\begin{align}
\mathcal{L}_{\tilde{\lambda}} \psi^\mu (\boldsymbol{x}) & \equiv \int L_{\tilde{\lambda}} (\boldsymbol{x}, \boldsymbol{x}') \psi^\mu(\boldsymbol{x}')d \boldsymbol{x}' \label{eq:pseudo-differential} \\
L_{\tilde{\lambda}} (\boldsymbol{x}, \boldsymbol{x}') & = \frac{1}{(2 \pi \mu)^d} \int_{\mathbb{R}^d} \tilde{\lambda} \left( \frac{\boldsymbol{x} + \boldsymbol{x}'}{2}, \boldsymbol{p} \right) e^{{\rm i} \boldsymbol{p} \cdot (\boldsymbol{x} - \boldsymbol{x}') / \mu} d \boldsymbol{p} \label{eq:kernel-function} \\
{\tilde{\lambda}} (\boldsymbol{x}, \boldsymbol{p}) & \equiv \begin{cases} \lambda(\boldsymbol{x}, \boldsymbol{p}) + \int_{\mathbb{R}^+} \beta(\omega) h(\omega - \lvert \boldsymbol{p} \rvert^2 / 2) d\omega \quad (\boldsymbol{x} \in \Gamma) \\
0 \quad (\boldsymbol{x} \notin \Gamma) . \label{eq:tilde-lambda}
\end{cases}
\end{align}
\end{subequations}
In (\ref{eq:microcanonical-exponent}) and (\ref{eq:pseudo-differential}), the integration range is unbounded. However, Riemann-Lebesgue lemma applied to (\ref{eq:kernel-function}) assures that the kernel function $L_{\tilde{\lambda}}$ vanishes in the limit of $\mu \to 0$ except in the vicinity of the points of $\boldsymbol{x} = \boldsymbol{x}'$. Consequently, further taking into account (\ref{eq:tilde-lambda}), the range of the integration of (\ref{eq:microcanonical-exponent}) and (\ref{eq:pseudo-differential}) can be reduced from $\mathbb{R}^d$ to $\Gamma$.

To obtain a simpler form of $f_G[\lambda, \beta]$, we need to compute the functional determinant of $\mathcal{L}_{\tilde{\lambda}}$. This is done here straightforwardly:
\begin{align}
f_G[\lambda, \beta] & = - \lim_{\mu \to 0} \left( 2 \pi \mu \right)^d \log \mathbb{E}_G \left\{ \exp \left[ - \frac{1}{\left( 2 \pi \mu \right)^d} \int_{\mathbb{R}^d} \psi^{\mu\dag} (\boldsymbol{x}) \mathcal{L}_{\tilde{\lambda}} \psi^\mu (\boldsymbol{x}) d\boldsymbol{x} \right] \right\} \nonumber \\
& = - \lim_{\mu \to 0} \lim_{\Delta \to 0} \left( 2 \pi \mu \right)^d \log \int \prod_{\boldsymbol{x} \in \Gamma_\Delta} \frac{d \psi^\mu(\boldsymbol{x}) d \psi^{\mu \dag}(\boldsymbol{x})}{2 \pi (2\pi\mu)^d \mathcal{N}} \nonumber \\
& \times \exp \left[ - \frac{\Delta^{2d}}{\left( 2 \pi \mu \right)^d} \sum_{\boldsymbol{x}, \boldsymbol{x}' \in \Gamma_\Delta} \psi^{\mu\dag} (\boldsymbol{x}) L_{\tilde{\lambda}} (\boldsymbol{x}, \boldsymbol{x}') \psi^\mu (\boldsymbol{x}') - \frac{\Delta^{d}}{\left( 2 \pi \mu \right)^d} \sum_{\boldsymbol{x} \in \Gamma_\Delta} \lvert \psi^\mu (\boldsymbol{x}) \rvert^2 \right] \nonumber \\
& = \lim_{\mu \to 0} \lim_{\Delta \to 0} \left( 2 \pi \mu \right)^d \log {\rm det} \left( \boldsymbol{I} + \boldsymbol{L}^\Delta_{\tilde{\lambda}} \right) .
\end{align}
We have carried out the Gaussian integration. The problem thus reduces to computing the determinant of an $\mathcal{N} \times \mathcal{N}$ matrix, $\boldsymbol{I} + \boldsymbol{L}^\Delta_{\tilde{\lambda}}$, where $\boldsymbol{I}$ is a unit matrix and $\boldsymbol{L}^\Delta_{\tilde{\lambda}}$ consists of $\left\{ \Delta^d L_{\tilde{\lambda}} (\boldsymbol{x}, \boldsymbol{x}') \ \vert \ \boldsymbol{x},\boldsymbol{x}' \in \Gamma_\Delta \right\}$. Let us use the following expression,
\begin{align}
\log {\rm det} \left( \boldsymbol{I} + \boldsymbol{L}^\Delta_{\tilde{\lambda}} \right)
& = {\rm tr} \log \left( \boldsymbol{I} + \boldsymbol{L}^\Delta_{\tilde{\lambda}} \right) \nonumber \\
& = - \sum_{j=1}^\infty \frac{(-1)^j}{j} {\rm tr} \boldsymbol{L}^{\Delta j}_{\tilde{\lambda}} ,
\end{align}
which holds for sufficiently small $\tilde{\lambda}$. We know that  $\boldsymbol{L}_{\tilde{\lambda}}^{\Delta j}$ consists of $\left\{ \Delta^d L^j_{\tilde{\lambda}} (\boldsymbol{x}, \boldsymbol{x}') \ \vert \ \boldsymbol{x},\boldsymbol{x}' \in \Gamma_\Delta \right\}$ where $L^j_{\tilde{\lambda}} (\boldsymbol{x}, \boldsymbol{x}')$ corresponds to the kernel function of an operator,
\begin{align}
\mathcal{L}^j_{\tilde{\lambda}} \equiv \underbrace{\mathcal{L}_{\tilde{\lambda}} \ldots \mathcal{L}_{\tilde{\lambda}}}_j ,
\end{align}
in the small $\Delta$ limit. In general, a product between pseudo-differential operators corresponds to a star product, or a Moyal product, between symbols \citep{polkovnikov2010phase,onuki2020quasi}. The star product is expanded in terms of $\mu$ with the leading term equivalent to the ordinary product. Accordingly, $\mathcal{L}^j_{\tilde{\lambda}} = \mathcal{L}_{\tilde{\lambda}^j} + \mathcal{O}(\mu)$ holds, where $\tilde{\lambda}^j$ is the $j$th power of $\tilde{\lambda}$. We thus derive
\begin{align}
\lim_{\Delta \to 0} L^j_{\tilde{\lambda}} (\boldsymbol{x}, \boldsymbol{x}) & = \frac{1}{(2 \pi \mu)^d} \int_{\mathbb{R}^d} \tilde{\lambda}^j (\boldsymbol{x}, \boldsymbol{p}) d\boldsymbol{p} + \mathcal{O}(\mu) \nonumber \\
\therefore \lim_{\Delta \to 0} {\rm tr} \boldsymbol{L}^{\Delta j}_{\tilde{\lambda}} & = \frac{1}{(2 \pi \mu)^d} \int_{\Gamma \times \mathbb{R}^d} \tilde{\lambda}^j (\boldsymbol{x}, \boldsymbol{p}) d\boldsymbol{x} d\boldsymbol{p} + \mathcal{O}(\mu) ,
\end{align}
and hence
\begin{align}
f_G[\lambda, \beta] = \int_{\Gamma \times \mathbb{R}^d} \log \left( \lambda(\boldsymbol{x}, \boldsymbol{p}) + \int_{\mathbb{R}^+} \beta(\omega) h \left(\omega - \frac{\lvert \boldsymbol{p} \rvert^2}{2} \right) d\omega + 1 \right) d\boldsymbol{x} d\boldsymbol{p} .
\end{align}
From this formula, the G\"artner-Ellis theorem yields the rate function $\mathcal{I}_G$ (\ref{eq:rate_G}) as
\begin{align} \label{eq:IG}
\mathcal{I}_G [n, A]=- \inf_{\lambda, \beta} \left\{\int d\boldsymbol{x} d\boldsymbol{p} \lambda(\boldsymbol{x}, \boldsymbol{p}) n(\boldsymbol{x}, \boldsymbol{p}) + \int_{\mathbb{R}^+} d \omega \beta(\omega) A(\omega) - f[\lambda, \beta] \right\} ,
\end{align}
which is computed as
\begin{align}
\mathcal{I}_G [n, A] = \begin{cases}
\int d\boldsymbol{x} d\boldsymbol{p} \left( n - 1 - \log n \right) & \mbox{if} \quad \mathcal{A}_\omega[n] = A(\omega)  \\
+ \infty & \mbox{otherwise}
\end{cases} .
\end{align}
As should have been expected, the minimum value of $\mathcal{I}_G[n, A]$ is $0$, which is realized when and only when $n = 1$ and $A(\omega) = \mathcal{A}_\omega[1]$.

Based on the large deviation result (\ref{eq:rate_G}), the numerator in (\ref{eq:microcanonical-distribution-2}) turns out to be
\begin{align} \label{eq:numerator}
 &\exp\left[ \frac{1}{(2\pi\mu)^{d}} \int_{\Gamma \times \mathbb{R}^d} n (\boldsymbol{x}, \boldsymbol{p}) d\boldsymbol{x} d\boldsymbol{p} \right] \mathbb{E}_G\left[ \delta \left( n^\mu - n \right) \Pi_\omega \delta \left( \mathcal{A}_\omega \left[ n^\mu \right] - A(\omega) \right) \right] \nonumber \\
\underset{\mu\to0}{\asymp} & \exp \left( \frac{S_A[n]}{(2 \pi \mu)^d} \right)
\end{align}
with
\begin{align} \label{eq:entropy}
S_A[n] = \begin{cases}
\int d\boldsymbol{x} d\boldsymbol{p} \left( 1 + \log n \right) & \mbox{if} \quad \mathcal{A}_\omega[n] = A(\omega)  \\
- \infty & \mbox{otherwise}
\end{cases} .
\end{align}
The finite part of this function defines the entropy for a mesoscopic state specified by $n$, and it coincides with the Lyapunov function \eqref{eq:quasipotential-unnorm} for the wave kinetic equation. Because the denominator of (\ref{eq:microcanonical-distribution-2}) is the integration of the numerator over all $n$, the Laplace's principle enables us to compute it as
\begin{align}
& \mathbb{E}_G \left\{ \exp\left[ \frac{1}{(2\pi\mu)^{d}} \int_{\Gamma \times \mathbb{R}^d} n^\mu (\boldsymbol{x}, \boldsymbol{p}) d\boldsymbol{x} d\boldsymbol{p} \right] \Pi_\omega \delta \left( \mathcal{A}_\omega \left[ n^\mu \right] - A(\omega) \right)  \right\} \nonumber \\
\underset{\mu\to0}{\asymp} & \exp \left(\frac{s[A]}{(2 \pi \mu)^d} \right),
\label{eq:denominator}
\end{align}
with $s[A] = \sup_n \left\{ S_A[n] \right\}$. The supremum is achieved when $\mathcal{A}_\omega[n] = A(\omega)$, and
\begin{align}
n(\boldsymbol{x}, \boldsymbol{p}) = n_h^A(\boldsymbol{p}) = N \left( \frac{\lvert \bp \rvert^2}{2} \right),
\end{align}
that is, $n$ is homogeneous in space and depends only on the magnitude of its wave vector. The function $N(\omega)$ is related to $A(\omega)$ by the condition $\mathcal{A}_\omega[n^A_h] = A(\omega)$. This gives formula (\ref{eq:stationary_WKE_2}) for $n_h^A$. We also have
\begin{align} \label{eq:s_A}
s[A] = \int d\boldsymbol{x} d\boldsymbol{p} \left( 1 + \log n_h^A \right) .
\end{align}
Finally, starting from (\ref{eq:microcanonical-distribution-2}), and using the two asymptotic relations (\ref{eq:numerator}) and (\ref{eq:denominator}), as well as \eqref{eq:entropy} and (\ref{eq:s_A}), we obtain
\begin{align}
\mathbb{P}^\mu_{A,m}[n^\mu = n] \underset{\mu\to0}{\asymp} \exp \left(-\frac{\mathcal{U}_A[n]}{(2 \pi \mu)^d} \right)
\end{align}
where the quasipotential $\mathcal{U}_A[n]$ is given by equation (\ref{eq:quasipotential-2}). We have established the announced results.

\section{Computations of scattering terms} \label{sec:scatter}
This appendix describes somewhat intricate derivation of the scattering terms that appear in the wave kinetic equation and the large deviation Hamiltonian. For this purpose, we prepare some useful formulae,
\begin{subequations}
\begin{align}
\int_0^t d\tau_1 e^{i \omega \tau_1 / \mu} \int_0^{\tau_1} d\tau_2 e^{-i \omega \tau_2 / \mu} & = \pi \mu t \delta(\omega) + o(\mu) \label{eq:formula1} \\
\int_0^t d\tau_1 e^{i \omega \tau_1 / \mu} \int_0^t d\tau_2 e^{-i \omega \tau_2 / \mu} & = 2 \pi \mu t \delta(\omega) + o(\mu) \label{eq:formula2} \\
\frac{1}{(2 \pi \mu)^d} \int_{\mathbb{R}^d} d \boldsymbol{\xi} e^{{\rm i} \boldsymbol{p} \cdot (\boldsymbol{\xi} - \boldsymbol{x}) / \mu} f(\boldsymbol{\xi}) & = \sum_{\lvert \alpha \rvert \geq 0} \frac{(- {\rm i} \mu)^{\lvert \alpha \rvert}}{\alpha !} \nabla_x^\alpha f(\boldsymbol{x}) \nabla_p^\alpha \delta (\boldsymbol{p}) . \label{eq:formula3}
\end{align}
\end{subequations}
Equations \eqref{eq:formula1} and \eqref{eq:formula2} are often used in literature of weak turbulence \cite{GBE}. The residual terms denoted by $o(\mu)$ make negligible contributions in the limit of $\mu \to 0$ compared to the leading-order terms when integrated with respect to $\omega$. In (\ref{eq:formula3}), a multi-index notation is used. In the following computation, integration is always carried out over $\mathbb{R}^d$, except for the basic positional coordinates represented by $\bx$ whose integration range is $\Gamma$.

\subsection{Terms appearing in the classical wave kinetic equation}
We first compute the scattering terms in the wave kinetic equation, specifically $\mathbb{E} \left[ w^\mu (\psi^\mu_2, \psi^\mu_0) \right]$, $\mathbb{E} \left[ w^\mu (\psi^\mu_0, \psi^\mu_2) \right]$, and $\mathbb{E} \left[ w^\mu (\psi^\mu_1, \psi^\mu_1) \right]$. These terms are common with those linear to $\lambda$ in the scattering part of the large deviation Hamiltonian, $\mathcal{H}_S$. The computations are slightly involved but mostly straightforward. A detailed procedure is presented only for the $\mathbb{E} \left[ w^\mu (\psi^\mu_2, \psi^\mu_0) \right]$ case.

From \eqref{eq:expansion_psi} and \eqref{eq:Wigner_transform}, we have
\begin{align*}
& w^\mu (\psi^\mu_2, \psi^\mu_0) = \frac{1}{- \mu^2 (2 \pi \mu)^d} \int_0^t d\tau_1 \int_0^{\tau_1} d\tau_2 \int d\boldsymbol{y} d\boldsymbol{\xi}_{1234} e^{- i \boldsymbol{p} \cdot \boldsymbol{y} / \mu} \\
\times & G^\mu \left( \boldsymbol{x} + \frac{\boldsymbol{y}}{2} - \boldsymbol{\xi}_1, t - \tau_1 \right) V^\mu(\boldsymbol{\xi}_1) G^\mu (\boldsymbol{\xi}_1 - \boldsymbol{\xi}_2, \tau_1 - \tau_2) V^\mu(\boldsymbol{\xi}_2) G^\mu (\boldsymbol{\xi}_2 - \boldsymbol{\xi}_3, \tau_2) \psi^\mu (\boldsymbol{\xi}_3, 0) \\
\times & G^{\mu\dag} \left( \boldsymbol{x} - \frac{\boldsymbol{y}}{2} - \boldsymbol{\xi}_4, t \right) \psi^{\mu\dag} (\boldsymbol{\xi}_4, 0) .
\end{align*}
Taking ensemble average, writing the propagators $G^\mu$ as Fourier integrals \eqref{eq:propagator} with wave vectors $\boldsymbol{\eta}_1, \boldsymbol{\eta}_2, \boldsymbol{\eta}_3$ and $\boldsymbol{\eta}_4$ in this order, and setting
\begin{align*}
\mathbb{E} \left[ V^\mu(\boldsymbol{\xi}_1) V^\mu(\boldsymbol{\xi}_2) \right] & = \int d \boldsymbol{\eta}_5 e^{i \boldsymbol{\eta}_5 \cdot (\boldsymbol{\xi}_1 - \boldsymbol{\xi}_2) / \mu} \Pi (\boldsymbol{\eta}_5) \\
\psi^\mu (\boldsymbol{\xi}_3, 0) \psi^{\mu\dag} (\boldsymbol{\xi}_4, 0) & = \int d \boldsymbol{\eta}_6 e^{i \boldsymbol{\eta}_6 \cdot (\boldsymbol{\xi}_3 - \boldsymbol{\xi}_4) / \mu} n ((\boldsymbol{\xi}_3 + \boldsymbol{\xi}_4)/2, \boldsymbol{\eta}_6) ,
\end{align*}
we derive
\begin{align*}
& \mathbb{E} \left[ w^\mu (\psi^\mu_2, \psi^\mu_0) \right] \\
= & \frac{1}{- \mu^2 (2 \pi \mu)^{5d}} \int d\boldsymbol{y} d\boldsymbol{\xi}_{1234} d\boldsymbol{\eta}_{123456} e^{- i \boldsymbol{p} \cdot \boldsymbol{y} / \mu} \\
\times & e^{-i (\vert\boldsymbol{\eta}_1\vert^2 - \vert\boldsymbol{\eta}_4\vert^2)t / 2 \mu} \int_0^t d\tau_1 e^{i (\vert\boldsymbol{\eta}_1\vert^2 - \vert\boldsymbol{\eta}_2\vert^2) \tau_1 / 2 \mu} \int_0^{\tau_1} d\tau_2 e^{i (\vert\boldsymbol{\eta}_2\vert^2 - \vert\boldsymbol{\eta}_3\vert^2) \tau_2 / 2 \mu} \\
\times & e^{i \boldsymbol{\eta}_1 \cdot (\boldsymbol{x} + \boldsymbol{y} / 2 - \boldsymbol{\xi}_1) / \mu} e^{i \boldsymbol{\eta}_2 \cdot (\boldsymbol{\xi}_1 - \boldsymbol{\xi}_2) / \mu} e^{i \boldsymbol{\eta}_3 \cdot (\boldsymbol{\xi}_2 - \boldsymbol{\xi}_3) / \mu} e^{- i \boldsymbol{\eta}_4 \cdot (\boldsymbol{x} - \boldsymbol{y} / 2 - \boldsymbol{\xi}_4) / \mu} \\
\times & \Pi (\boldsymbol{\eta}_5) e^{i \boldsymbol{\eta}_5 \cdot (\boldsymbol{\xi}_1 - \boldsymbol{\xi}_2)} n((\boldsymbol{\xi}_3 + \boldsymbol{\xi}_4) / 2, \boldsymbol{\eta}_6) e^{i \boldsymbol{\eta}_6 \cdot (\boldsymbol{\xi}_3 - \boldsymbol{\xi}_4) / \mu} .
\end{align*}
Integration of this expression with respect to $\boldsymbol{y}$, $\boldsymbol{\xi}_1$ and $\boldsymbol{\xi}_2$ yields
\begin{align*}
& \mathbb{E} \left[ w^\mu (\psi^\mu_2, \psi^\mu_0) \right] \\\
= & \frac{1}{- \mu^2 (2 \pi \mu)^{2d}} \int d\boldsymbol{\xi}_{34} d\boldsymbol{\eta}_{123456} \\
\times & e^{-i (\vert\boldsymbol{\eta}_1\vert^2 - \vert\boldsymbol{\eta}_4\vert^2)t / 2 \mu} \int_0^t d\tau_1 e^{i (\vert\boldsymbol{\eta}_1\vert^2 - \vert\boldsymbol{\eta}_2\vert^2) \tau_1 / 2 \mu} \int_0^{\tau_1} d\tau_2 e^{i (\vert\boldsymbol{\eta}_2\vert^2 - \vert\boldsymbol{\eta}_3\vert^2) \tau_2 / 2 \mu} \\
\times & e^{i (\boldsymbol{\eta}_1 \cdot \boldsymbol{x} - \boldsymbol{\eta}_3 \cdot \boldsymbol{\xi}_3 - \boldsymbol{\eta}_4 \cdot \boldsymbol{x} + \boldsymbol{\eta}_4 \cdot \boldsymbol{\xi}_4 + \boldsymbol{\eta}_6 \cdot \boldsymbol{\xi}_3 - \boldsymbol{\eta}_6 \cdot \boldsymbol{\xi}_4) / \mu} \\
\times & \delta((\boldsymbol{\eta}_1 + \boldsymbol{\eta}_4) / 2 - \boldsymbol{p}) \delta(\boldsymbol{\eta}_1 - \boldsymbol{\eta}_2 - \boldsymbol{\eta}_5) \delta(\boldsymbol{\eta}_3 - \boldsymbol{\eta}_2 - \boldsymbol{\eta}_5) \\
\times & \Pi (\boldsymbol{\eta}_5) n((\boldsymbol{\xi}_3 + \boldsymbol{\xi}_4) / 2, \boldsymbol{\eta}_6) .
\end{align*}
We change the variables as
\begin{align*}
\boldsymbol{X} = \frac{\boldsymbol{\xi}_3 + \boldsymbol{\xi}_4}{2}, \quad \boldsymbol{Y} = \boldsymbol{\xi}_3 - \boldsymbol{\xi}_4 ,
\end{align*}
and carry out the integration with respect to $\boldsymbol{Y}$ to get
\begin{align*}
& \mathbb{E} \left[ w^\mu (\psi^\mu_2, \psi^\mu_0) \right] \\
= & \frac{1}{- \mu^2 (2 \pi \mu)^d} \int d\boldsymbol{X} d\boldsymbol{\eta}_{123456} \\
\times & e^{-i (\vert\boldsymbol{\eta}_1\vert^2 - \vert\boldsymbol{\eta}_4\vert^2)t / 2 \mu} \int_0^t d\tau_1 e^{i (\vert\boldsymbol{\eta}_1\vert^2 - \vert\boldsymbol{\eta}_2\vert^2) \tau_1 / 2 \mu} \int_0^{\tau_1} d\tau_2 e^{i (\vert\boldsymbol{\eta}_2\vert^2 - \vert\boldsymbol{\eta}_3\vert^2) \tau_2 / 2 \mu} \\
\times & e^{i (\boldsymbol{\eta}_1 - \boldsymbol{\eta}_4) \cdot (\boldsymbol{x} - \boldsymbol{X}) / \mu} \delta((\boldsymbol{\eta}_3 + \boldsymbol{\eta}_4) / 2 - \boldsymbol{\eta}_6) \\
\times & \delta((\boldsymbol{\eta}_1 + \boldsymbol{\eta}_4) / 2 - \boldsymbol{p}) \delta(\boldsymbol{\eta}_1 - \boldsymbol{\eta}_2 - \boldsymbol{\eta}_5) \delta(\boldsymbol{\eta}_3 - \boldsymbol{\eta}_2 - \boldsymbol{\eta}_5) \\
\times & \Pi (\boldsymbol{\eta}_5) n(\boldsymbol{X}, \boldsymbol{\eta}_6) .
\end{align*}
We understand that $\boldsymbol{\eta}_1 = \boldsymbol{\eta}_3$ and $\boldsymbol{\eta}_6 = \boldsymbol{p}$ hold in the integrand. Integration with respect to $\boldsymbol{\eta}_3$ and $\boldsymbol{\eta}_6$ yields
\begin{align*}
& \mathbb{E} \left[ w^\mu (\psi^\mu_2, \psi^\mu_0) \right] \\
= & \frac{1}{- \mu^2 (2 \pi \mu)^{d}} \int d\boldsymbol{X} d\boldsymbol{\eta}_{1245} \\
\times & e^{-i (\vert\boldsymbol{\eta}_1\vert^2 - \vert\boldsymbol{\eta}_4\vert^2)t / 2 \mu} \int_0^t d\tau_1 e^{i (\vert\boldsymbol{\eta}_1\vert^2 - \vert\boldsymbol{\eta}_2\vert^2) \tau_1 / 2 \mu} \int_0^{\tau_1} d\tau_2 e^{i (\vert\boldsymbol{\eta}_2\vert^2 - \vert\boldsymbol{\eta}_1\vert^2) \tau_2 / 2 \mu} \\
\times & e^{i (\boldsymbol{\eta}_1 - \boldsymbol{\eta}_4) \cdot (\boldsymbol{x} - \boldsymbol{X}) / \mu} \delta((\boldsymbol{\eta}_1 + \boldsymbol{\eta}_4) / 2 - \boldsymbol{p}) \delta(\boldsymbol{\eta}_1 - \boldsymbol{\eta}_2 - \boldsymbol{\eta}_5) \\
\times & \Pi (\boldsymbol{\eta}_5) n(\boldsymbol{X}, \boldsymbol{p}) .
\end{align*}
We use (\ref{eq:formula1}) to derive
\begin{align*}
& \int_0^t d\tau_1 e^{i (\vert\boldsymbol{\eta}_1\vert^2 - \vert\boldsymbol{\eta}_2\vert^2) \tau_1 / 2 \mu} \int_0^{\tau_1} d\tau_2 e^{i (\vert\boldsymbol{\eta}_2\vert^2 - \vert\boldsymbol{\eta}_1\vert^2) \tau_2 / 2 \mu} \\
= & \pi \mu t \delta \left( \frac{\vert\boldsymbol{\eta}_1\vert^2}{2} - \frac{\vert\boldsymbol{\eta}_2\vert^2}{2} \right) + o(\mu) .
\end{align*}
We also use (\ref{eq:formula3}) to derive
\begin{align*}
& \frac{1}{(2 \pi \mu)^{d}} \int d\boldsymbol{X} d\boldsymbol{\eta}_{4} e^{-i (\vert\boldsymbol{\eta}_1\vert^2 - \vert\boldsymbol{\eta}_4\vert^2)t / 2 \mu} e^{i (\boldsymbol{\eta}_1 - \boldsymbol{\eta}_4) \cdot (\boldsymbol{x} - \boldsymbol{X}) / \mu} \delta((\boldsymbol{\eta}_1 + \boldsymbol{\eta}_4) / 2 - \boldsymbol{p}) n (\boldsymbol{X}, \boldsymbol{p}) \\
= & \delta(\boldsymbol{\eta}_1 - \boldsymbol{p}) n(\boldsymbol{x}, \boldsymbol{p}) + \mathcal{O} (\mu, t) .
\end{align*}
Consequently, we obtain
\begin{align} \label{eq:term1-1}
\mathbb{E} \left[ w^\mu (\psi^\mu_2, \psi^\mu_0) \right] & = - \frac{t}{2 \mu} \int d\boldsymbol{\eta} \sigma (\boldsymbol{p}, \boldsymbol{\eta}) n (\boldsymbol{x}, \boldsymbol{p}) + o \left( \frac{t}{\mu} \right) .
\end{align}
Because $w^\mu (\psi^\mu_0, \psi^\mu_2) = [w^\mu (\psi^\mu_2, \psi^\mu_0)]^\dag$, and $\sigma$ and $n$ are real functions, we also have
\begin{align} \label{eq:term1-2}
\mathbb{E} \left[ w^\mu (\psi^\mu_0, \psi^\mu_2) \right] & = - \frac{t}{2 \mu} \int d\boldsymbol{\eta} \sigma (\boldsymbol{p}, \boldsymbol{\eta}) n (\boldsymbol{x}, \boldsymbol{p}) + o \left( \frac{t}{\mu} \right) .
\end{align}
Finally, we consider $\mathbb{E}[w^\mu (\psi^\mu_1, \psi^\mu_1)]$. From \eqref{eq:expansion_psi} and \eqref{eq:Wigner_transform}, we have
\begin{align*}
w^\mu (\psi^\mu_1, \psi^\mu_1) & = \frac{1}{\mu^2 (2 \pi \mu)^d} \int_0^t d\tau_1 \int_0^t d\tau_2 \int d\boldsymbol{y} d\boldsymbol{\xi}_{1234} e^{- i \boldsymbol{p} \cdot \boldsymbol{y} / \mu}\\
& \times G^\mu \left( \boldsymbol{x} + \frac{\boldsymbol{y}}{2} - \boldsymbol{\xi}_1, t - \tau_1 \right) V^\mu(\boldsymbol{\xi}_1) G^\mu (\boldsymbol{\xi}_1 - \boldsymbol{\xi}_2, \tau_1) \psi^\mu (\boldsymbol{\xi}_2, 0) \\
& \times G^{\mu\dag} \left( \boldsymbol{x} - \frac{\boldsymbol{y}}{2} - \boldsymbol{\xi}_3, t - \tau_2 \right) V^\mu(\boldsymbol{\xi}_3) G^{\mu\dag} (\boldsymbol{\xi}_3 - \boldsymbol{\xi}_4, \tau_2) \psi^{\mu\dag} (\boldsymbol{\xi}_4, 0) .
\end{align*}
Taking the ensemble average, introducing the Fourier integrals, and integrating some variables in the same manner as the previous case, we derive
\begin{align*}
& \mathbb{E} \left[ w^\mu (\psi^\mu_1, \psi^\mu_1) \right] \\
= & \frac{1}{\mu^2 (2 \pi \mu)^d} \int d\boldsymbol{X} d\boldsymbol{\eta}_{123456} \\
\times & e^{-i (\vert\boldsymbol{\eta}_1\vert^2 - \vert\boldsymbol{\eta}_3\vert^2)t / 2 \mu} \int_0^t d\tau_1 e^{i (\vert\boldsymbol{\eta}_1\vert^2 - \vert\boldsymbol{\eta}_2\vert^2) \tau_1 / 2 \mu} \int_0^t d\tau_2 e^{- i (\vert\boldsymbol{\eta}_3\vert^3 - \vert\boldsymbol{\eta}_4\vert^2) \tau_2 / 2 \mu} \\
\times & e^{i (\boldsymbol{\eta}_1 - \boldsymbol{\eta}_3) \cdot (\boldsymbol{x} - \boldsymbol{X}) / \mu} \delta( (\boldsymbol{\eta}_2 + \boldsymbol{\eta}_4) / 2 - \boldsymbol{\eta}_6) \\
\times & \delta((\boldsymbol{\eta}_1 + \boldsymbol{\eta}_3) / 2 - \boldsymbol{p}) \delta(\boldsymbol{\eta}_1 - \boldsymbol{\eta}_2 - \boldsymbol{\eta}_5) \delta(\boldsymbol{\eta}_3 - \boldsymbol{\eta}_4 - \boldsymbol{\eta}_5) \\
\times & \Pi (\boldsymbol{\eta}_5) n(\boldsymbol{X}, \boldsymbol{\eta}_6) .
\end{align*}
Applying (\ref{eq:formula3}), understanding $\boldsymbol{\eta}_1 = \boldsymbol{\eta}_3$ and $\boldsymbol{\eta}_2 = \boldsymbol{\eta}_4$ in the integrand, using (\ref{eq:formula2}), and Integrating all the possible variables, we finally obtain
\begin{align} \label{eq:term1-3}
\mathbb{E} \left[ w^\mu (\psi^\mu_1, \psi^\mu_1) \right] = \frac{t}{\mu} \int d\boldsymbol{\eta} \sigma(\boldsymbol{p}, \boldsymbol{\eta}) n(\boldsymbol{x}, \boldsymbol{\eta}) + o \left( \frac{t}{\mu} \right) .
\end{align}

\subsection{Quadratic terms in the Hamiltonian}
We compute the terms in $\mathcal{H}_S$ quadratic in $\lambda$, originating from four expressions,
\begin{subequations}
\begin{align}
\frac{1}{(2\pi\mu)^{2d}} \int d\boldsymbol{x}_{12} d\boldsymbol{p}_{12} \lambda(\boldsymbol{x}_1, \boldsymbol{p}_1) \lambda(\boldsymbol{x}_2, \boldsymbol{p}_2) \mathbb{E}[w^\mu(\psi^\mu_1, \psi^\mu_0)(\boldsymbol{x}_1, \boldsymbol{p}_1) w^\mu(\psi^\mu_0, \psi^\mu_1)(\boldsymbol{x}_2, \boldsymbol{p}_2)] \label{eq:quadratic1} \\
\frac{1}{(2\pi\mu)^{2d}} \int d\boldsymbol{x}_{12} d\boldsymbol{p}_{12} \lambda(\boldsymbol{x}_1, \boldsymbol{p}_1) \lambda(\boldsymbol{x}_2, \boldsymbol{p}_2) \mathbb{E}[w^\mu(\psi^\mu_0, \psi^\mu_1)(\boldsymbol{x}_1, \boldsymbol{p}_1) w^\mu(\psi^\mu_1, \psi^\mu_0)(\boldsymbol{x}_2, \boldsymbol{p}_2)] \label{eq:quadratic2} \\
\frac{1}{(2\pi\mu)^{2d}} \int d\boldsymbol{x}_{12} d\boldsymbol{p}_{12} \lambda(\boldsymbol{x}_1, \boldsymbol{p}_1) \lambda(\boldsymbol{x}_2, \boldsymbol{p}_2) \mathbb{E}[w^\mu(\psi^\mu_1, \psi^\mu_0)(\boldsymbol{x}_1, \boldsymbol{p}_1) w^\mu(\psi^\mu_1, \psi^\mu_0)(\boldsymbol{x}_2, \boldsymbol{p}_2)] \label{eq:quadratic3} \\
\frac{1}{(2\pi\mu)^{2d}} \int d\boldsymbol{x}_{12} d\boldsymbol{p}_{12} \lambda(\boldsymbol{x}_1, \boldsymbol{p}_1) \lambda(\boldsymbol{x}_2, \boldsymbol{p}_2) \mathbb{E}[w^\mu(\psi^\mu_0, \psi^\mu_1)(\boldsymbol{x}_1, \boldsymbol{p}_1) w^\mu(\psi^\mu_0, \psi^\mu_1)(\boldsymbol{x}_2, \boldsymbol{p}_2)] . \label{eq:quadratic4}
\end{align}
\end{subequations}
Among these, the two pairs, \eqref{eq:quadratic1}-\eqref{eq:quadratic2} and \eqref{eq:quadratic3}-\eqref{eq:quadratic4}, are complex conjugate, respectively. Therefore, we need to compute only two expressions. Although the number of factors involved in the integration is greater than those in the linear terms, the computation procedures are largely the same. For the sake of conciseness, we denote time $t$ instead of $\Delta t$.

From \eqref{eq:expansion_psi} and \eqref{eq:Wigner_transform}, we write the Wigner transforms in the integrand of \eqref{eq:quadratic1} as
\begin{align*}
w^\mu(\psi^\mu_1, \psi^\mu_0) (\boldsymbol{x}_1, \boldsymbol{p}_1) = & \frac{1}{i \mu (2 \pi \mu)^d} \int_0^t d\tau d \boldsymbol{y} d\boldsymbol{\xi}_{123} e^{- i \boldsymbol{p}_1 \cdot \boldsymbol{y} / \mu} \\
\times & G^\mu \left( \boldsymbol{x}_1 + \frac{\boldsymbol{y}}{2} - \boldsymbol{\xi}_1, t - \tau \right) V^\mu (\boldsymbol{\xi}_1) G^\mu (\boldsymbol{\xi}_1 - \boldsymbol{\xi}_2, \tau) \psi^\mu (\boldsymbol{\xi}_2, 0) \\
\times & G^{\mu \dag} \left( \boldsymbol{x}_1 - \frac{\boldsymbol{y}}{2} - \boldsymbol{\xi}_3, t \right) \psi^{\mu \dag} (\boldsymbol{\xi}_3, 0) \\
w^\mu(\psi^\mu_0, \psi^\mu_1) (\boldsymbol{x}_2, \boldsymbol{p}_2) = & \frac{1}{- i \mu (2 \pi \mu)^d} \int_0^t d\tau d \boldsymbol{y} d\boldsymbol{\xi}_{123} e^{- i \boldsymbol{p}_2 \cdot \boldsymbol{y} / \mu} \\
\times & G^\mu \left( \boldsymbol{x}_2 + \frac{\boldsymbol{y}}{2} - \boldsymbol{\xi}_1, t \right) \psi^\mu (\boldsymbol{\xi}_1, 0) \\
\times & G^{\mu \dag} \left( \boldsymbol{x}_2 - \frac{\boldsymbol{y}}{2} - \boldsymbol{\xi}_2, t - \tau \right) V^\mu (\boldsymbol{\xi}_2) G^{\mu \dag} (\boldsymbol{\xi}_2 - \boldsymbol{\xi}_3, \tau) \psi^{\mu \dag} (\boldsymbol{\xi}_3, 0) .
\end{align*}
Taking the ensemble average of the product of these expressions, it would be straightforward to have
\begin{align*}
& \frac{1}{(2\pi\mu)^{2d}} \int d\boldsymbol{x}_{12} d\boldsymbol{p}_{12} \lambda(\boldsymbol{x}_1, \boldsymbol{p}_1) \lambda(\boldsymbol{x}_2, \boldsymbol{p}_2) \mathbb{E}[w^\mu(\psi^\mu_1, \psi^\mu_0)(\boldsymbol{x}_1, \boldsymbol{p}_1) w^\mu(\psi^\mu_0, \psi^\mu_1)(\boldsymbol{x}_2, \boldsymbol{p}_2)] \\
= & \frac{1}{\mu^2 (2 \pi \mu)^{4d}} \int d\boldsymbol{x}_{12} d\boldsymbol{p}_{12} d\boldsymbol{X}_{12} d\boldsymbol{\eta}_{123456789} \lambda(\boldsymbol{x}_1, \boldsymbol{p}_1) \lambda(\boldsymbol{x}_2, \boldsymbol{p}_2) \\
\times & e^{-i (\vert\boldsymbol{\eta}_1\vert^2 - \vert\boldsymbol{\eta}_3\vert^2)t / 2 \mu} \int_0^t d\tau_1 e^{i (\vert\boldsymbol{\eta}_1\vert^2 - \vert\boldsymbol{\eta}_2\vert^2)\tau_1 / 2 \mu} \\
\times & e^{-i (\vert\boldsymbol{\eta}_4\vert^2 - \vert\boldsymbol{\eta}_5\vert^2)t / 2 \mu} \int_0^t d\tau_2 e^{- i (\vert\boldsymbol{\eta}_5\vert^2 - \vert\boldsymbol{\eta}_6\vert^2)\tau_2 / 2 \mu} \\
\times & \delta(\boldsymbol{\eta}_1 / 2 + \boldsymbol{\eta}_3 / 2 - \boldsymbol{p}_1) \delta(\boldsymbol{\eta}_4 / 2 + \boldsymbol{\eta}_5 / 2 - \boldsymbol{p}_2) \delta(\boldsymbol{\eta}_1  - \boldsymbol{\eta}_2 - \boldsymbol{\eta}_7) \delta(\boldsymbol{\eta}_5  - \boldsymbol{\eta}_6 - \boldsymbol{\eta}_7) \\
\times & \delta(\boldsymbol{\eta}_2 / 2 + \boldsymbol{\eta}_6 / 2 - \boldsymbol{\eta}_8) \delta(\boldsymbol{\eta}_3 / 2 + \boldsymbol{\eta}_4 / 2 - \boldsymbol{\eta}_9) \\
\times & e^{i (\boldsymbol{\eta}_1 - \boldsymbol{\eta}_3) \cdot \boldsymbol{x}_1 / \mu} e^{i (\boldsymbol{\eta}_4 - \boldsymbol{\eta}_5) \cdot \boldsymbol{x}_2 / \mu} e^{- i (\boldsymbol{\eta}_2 - \boldsymbol{\eta}_6) \cdot \boldsymbol{X}_1 / \mu} e^{- i (\boldsymbol{\eta}_4 - \boldsymbol{\eta}_3) \cdot \boldsymbol{X}_2 / \mu} \\
\times & \Pi (\boldsymbol{\eta}_7) n(\boldsymbol{X}_1, \boldsymbol{\eta}_8) n(\boldsymbol{X}_2, \boldsymbol{\eta}_9) .
\end{align*}
Applying (\ref{eq:formula3}) and (\ref{eq:formula1}) and integrating all the possible variables yields
\begin{align} \label{eq:term2-1}
& \frac{1}{(2\pi\mu)^{2d}} \int d\boldsymbol{x}_{12} d\boldsymbol{p}_{12} \lambda(\boldsymbol{x}_1, \boldsymbol{p}_1) \lambda(\boldsymbol{x}_2, \boldsymbol{p}_2) \mathbb{E}[w^\mu(\psi^\mu_1, \psi^\mu_0)(\boldsymbol{x}_1, \boldsymbol{p}_1) w^\mu(\psi^\mu_0, \psi^\mu_1)(\boldsymbol{x}_2, \boldsymbol{p}_2)] \nonumber \\
= & \frac{t}{\mu (2\pi\mu)^d} \int d\boldsymbol{x}_1 d\boldsymbol{p}_1 d\boldsymbol{\eta}_2 \lambda(\boldsymbol{x}_1, \boldsymbol{p}_1)^2 \sigma(\boldsymbol{p}_1, \boldsymbol{\eta}_2) n(\boldsymbol{x}_1, \boldsymbol{p}_1) n(\boldsymbol{x}_1, \boldsymbol{\eta}_2) + o \left( \frac{t}{\mu (2\pi\mu)^d} \right) .
\end{align}
From the the condition of complex conjugate, we also have
\begin{align} \label{eq:term2-2}
& \frac{1}{(2\pi\mu)^{2d}} \int d\boldsymbol{x}_{12} d\boldsymbol{p}_{12} \lambda(\boldsymbol{x}_1, \boldsymbol{p}_1) \lambda(\boldsymbol{x}_2, \boldsymbol{p}_2) \mathbb{E}[w^\mu(\psi^\mu_0, \psi^\mu_1)(\boldsymbol{x}_1, \boldsymbol{p}_1) w^\mu(\psi^\mu_1, \psi^\mu_0)(\boldsymbol{x}_2, \boldsymbol{p}_2)] \nonumber \\
= & \frac{t}{\mu (2\pi\mu)^d} \int d\boldsymbol{x}_1 d\boldsymbol{p}_1 d\boldsymbol{\eta}_2 \lambda(\boldsymbol{x}_1, \boldsymbol{p}_1)^2 \sigma(\boldsymbol{p}_1, \boldsymbol{\eta}_2) n(\boldsymbol{x}_1, \boldsymbol{p}_1) n(\boldsymbol{x}_1, \boldsymbol{\eta}_2) + o \left( \frac{t}{\mu (2\pi\mu)^d} \right) .
\end{align}

For the computation of \eqref{eq:quadratic3}, we write,
\begin{align*}
w^\mu(\psi^\mu_1, \psi^\mu_0) (\boldsymbol{x}_1, \boldsymbol{p}_1) = & \frac{1}{i \mu (2 \pi \mu)^d} \int_0^t d\tau d \boldsymbol{y} d\boldsymbol{\xi}_{123} e^{- i \boldsymbol{p}_1 \cdot \boldsymbol{y} / \mu} \\
\times & G^\mu \left( \boldsymbol{x}_1 + \frac{\boldsymbol{y}}{2} - \boldsymbol{\xi}_1, t - \tau \right) V^\mu (\boldsymbol{\xi}_1) G^\mu (\boldsymbol{\xi}_1 - \boldsymbol{\xi}_2, \tau) \psi^\mu (\boldsymbol{\xi}_2, 0) \\
\times & G^{\mu \dag} \left( \boldsymbol{x}_1 - \frac{\boldsymbol{y}}{2} - \boldsymbol{\xi}_3, t \right) \psi^{\mu \dag} (\boldsymbol{\xi}_3, 0) \\
\end{align*}
Taking the ensemble average of the product of this expression, it would be straightforward to have
\begin{align*}
& \frac{1}{(2\pi\mu)^{2d}} \int d\boldsymbol{x}_{12} d\boldsymbol{p}_{12} \lambda(\boldsymbol{x}_1, \boldsymbol{p}_1) \lambda(\boldsymbol{x}_2, \boldsymbol{p}_2) \mathbb{E}[w^\mu(\psi^\mu_1, \psi^\mu_0)(\boldsymbol{x}_1, \boldsymbol{p}_1) w^\mu(\psi^\mu_1, \psi^\mu_0)(\boldsymbol{x}_2, \boldsymbol{p}_2)] \\
= & \frac{1}{- \mu^2 (2 \pi \mu)^{4d}} \int d\boldsymbol{x}_{12} d\boldsymbol{p}_{12} d\boldsymbol{X}_{12} d\boldsymbol{\eta}_{123456789} \lambda(\boldsymbol{x}_1, \boldsymbol{p}_1) \lambda(\boldsymbol{x}_2, \boldsymbol{p}_2) \\
\times & e^{-i (\vert\boldsymbol{\eta}_1\vert^2 - \vert\boldsymbol{\eta}_3\vert^2)t / 2 \mu} \int_0^t d\tau_1 e^{i (\vert\boldsymbol{\eta}_1\vert^2 - \vert\boldsymbol{\eta}_2\vert^2)\tau_1 / 2 \mu} \\
\times & e^{-i (\vert\boldsymbol{\eta}_4\vert^2 - \vert\boldsymbol{\eta}_6\vert^2)t / 2 \mu} \int_0^t d\tau_2 e^{i (\vert\boldsymbol{\eta}_4\vert^2 - \vert\boldsymbol{\eta}_5\vert^2)\tau_2 / 2 \mu} \\
\times & \delta(\boldsymbol{\eta}_1 / 2 + \boldsymbol{\eta}_3 / 2 - \boldsymbol{p}_1) \delta(\boldsymbol{\eta}_4 / 2 + \boldsymbol{\eta}_6 / 2 - \boldsymbol{p}_2) \delta(\boldsymbol{\eta}_1  - \boldsymbol{\eta}_2 - \boldsymbol{\eta}_7) \delta(\boldsymbol{\eta}_4  - \boldsymbol{\eta}_5 + \boldsymbol{\eta}_7) \\
\times & \delta(\boldsymbol{\eta}_2 / 2 + \boldsymbol{\eta}_6 / 2 - \boldsymbol{\eta}_8) \delta(\boldsymbol{\eta}_5 / 2 + \boldsymbol{\eta}_3 / 2 - \boldsymbol{\eta}_9) \\
\times & e^{i (\boldsymbol{\eta}_1 - \boldsymbol{\eta}_3) \cdot \boldsymbol{x}_1 / \mu} e^{i (\boldsymbol{\eta}_4 - \boldsymbol{\eta}_6) \cdot \boldsymbol{x}_2 / \mu} e^{- i (\boldsymbol{\eta}_2 - \boldsymbol{\eta}_6) \cdot \boldsymbol{X}_1 / \mu} e^{- i (\boldsymbol{\eta}_5 - \boldsymbol{\eta}_3) \cdot \boldsymbol{X}_2 / \mu} \\
\times & \Pi (\boldsymbol{\eta}_7) n(\boldsymbol{X}_1, \boldsymbol{\eta}_8) n(\boldsymbol{X}_2, \boldsymbol{\eta}_9) .
\end{align*}
Applying (\ref{eq:formula3}) and (\ref{eq:formula1}) and integrating all the possible variables yields
\begin{align} \label{eq:term2-3}
& \frac{1}{(2\pi\mu)^{2d}} \int d\boldsymbol{x}_{12} d\boldsymbol{p}_{12} \lambda(\boldsymbol{x}_1, \boldsymbol{p}_1) \lambda(\boldsymbol{x}_2, \boldsymbol{p}_2) \mathbb{E}[w^\mu(\psi^\mu_1, \psi^\mu_0)(\boldsymbol{x}_1, \boldsymbol{p}_1) w^\mu(\psi^\mu_1, \psi^\mu_0)(\boldsymbol{x}_2, \boldsymbol{p}_2)] \nonumber \\
= & - \frac{t}{\mu (2\pi\mu)^d} \int d\boldsymbol{x}_1 d\boldsymbol{p}_{12} \lambda(\boldsymbol{x}_1, \boldsymbol{p}_1) \lambda(\boldsymbol{x}_1, \boldsymbol{p}_2) \sigma(\boldsymbol{p}_1, \boldsymbol{p}_2) n(\boldsymbol{x}_1, \boldsymbol{p}_1) n(\boldsymbol{x}_1, \boldsymbol{p}_2) + o \left( \frac{t}{\mu (2\pi\mu)^d} \right) .
\end{align}
Finally, from the the condition of complex conjugate, we have
\begin{align} \label{eq:term2-4}
& \frac{1}{(2\pi\mu)^{2d}} \int d\boldsymbol{x}_{12} d\boldsymbol{p}_{12} \lambda(\boldsymbol{x}_1, \boldsymbol{p}_1) \lambda(\boldsymbol{x}_2, \boldsymbol{p}_2) \mathbb{E}[w^\mu(\psi^\mu_0, \psi^\mu_1)(\boldsymbol{x}_1, \boldsymbol{p}_1) w^\mu(\psi^\mu_0, \psi^\mu_1)(\boldsymbol{x}_2, \boldsymbol{p}_2)] \nonumber \\
= & - \frac{t}{\mu (2\pi\mu)^d} \int d\boldsymbol{x}_1 d\boldsymbol{p}_{12} \lambda(\boldsymbol{x}_1, \boldsymbol{p}_1) \lambda(\boldsymbol{x}_1, \boldsymbol{p}_2) \sigma(\boldsymbol{p}_1, \boldsymbol{p}_2) n(\boldsymbol{x}_1, \boldsymbol{p}_1) n(\boldsymbol{x}_1, \boldsymbol{p}_2) + o \left( \frac{t}{\mu (2\pi\mu)^d} \right) .
\end{align}

\section{Derivation of the Hamiltonian in the diffusive limit}\label{app:diffusive_limit}
In this appendix, we show how to obtain the large deviation Hamiltonian \eqref{eq:large_dev_H_diffusive_limit} from \eqref{eq:scattering_Hamiltonian_DB}. Since we are dealing with the scattering term that is local in position space, one can omit the dependence on $\bx$. As explained in the main text, the important point is to show how the diffusion kernel $\Sigma[n]$ transforms when only infinitesimal deviation from incoming wave vector \yo{is} allowed by the cross section. For this specific purpose, we introduce a parameter $\nu$ and write the typical correlation length of $V$ as $\mu / \nu$. Our strategy is to expand the Hamiltonian in terms of $\nu$.

Because the potential spectrum $\Pi(\bp)$ is supposed to have a finite support with the extent of $\mathcal{O}(\nu)$, it is convenient to rewrite the cross section as $\sigma(\bp_1,\bp_2) = \nu^{-d}\tilde{\sigma}((\bp_1-\bp_2) / \nu; \bp_1)$. For any test functions, $f$ and $g$, the diffusive kernel \eqref{eq:def:operator_Sigma} satisfies
\begin{align}
& \int d \bp_1 d \bp_2 f(\bp_1)\Sigma(\bp_1,\bp_2)g(\bp_2) \nonumber \\
= & c \int d \bp d \bq f(\bp)n(\bp)\tilde{\sigma}(\bq;\bp)n(\bp - \nu \bq)\left[g(\bp) - g(\bp - \nu \bq)\right] .
\end{align}
Taylor expanding the integrand with respect to $\nu$, one obtains
\begin{align} \label{eq:diffusion_1}
& \int d \bp_1 d \bp_2 f(\bp_1)\Sigma(\bp_1, \bp_2)g(\bp_2) \nonumber \\
= & c \int d \bp f(\bp)n(\bp) \Biggl\{\nu n(\bp) \sum_{n=1}^{d} \left( \int dq_n \tilde{\sigma}(\bq; \bp) q_n\right) \partial_{p_n}g(\bp) \nonumber \\
- & \nu^2 \sum_{n=1}^{d}\sum_{m=1}^{d} \left( \int dq_n dq_m \tilde{\sigma}(\bq; \bp) q_n q_m \right)\partial_{p_n}n(\bp) \partial_{p_{m}}g(\bp) \nonumber \\
- & \frac{\nu^2}{2} n(\bp) \sum_{n=1}^{d}\sum_{m=1}^{d} \left(\int dq_n dq_m \tilde{\sigma}(\bq; \bp) q_n q_m \right) \partial_{p_n}\partial_{p_m} g(\bp) + \mathcal{O} \left( \nu^3 \right) \Biggr\} .
\end{align}
To evaluate the integrations with respect to $\boldsymbol{q}$, we shall exponentiate the Dirac-$\delta$ in the definition of the cross section (\ref{eq:def:cross_section}) and use the inverse Fourier transform of $\Pi$ in (\ref{eq:def:diffusion_matrix}). The resulting expression is
\begin{align} \label{eq:cross_section_tilde}
& \tilde{\sigma}(\bq ; \bp) = \frac{1}{\nu (2\pi)^{d}} \int d\by \int_{\mathbb{R}} ds \mathrm{e}^{- i\bq \cdot \by} R^\nu \left( s\bp + \by -\frac{\nu}{2}s\bq \right) \\
& \text{and} \quad R^\nu (\by) = \int d \bp \Pi(\bp) \mathrm{e}^{i\bp\cdot\by / \nu} , \nonumber
\end{align}
where $R^\nu(\boldsymbol{y}) = R(\boldsymbol{y} / \nu)$ (see Eq. (\ref{eq:def:diffusion_matrix})) is a scaled correlation function of the random potential with the correlation length of $\mathcal{O}(1)$. Again Taylor expanding Eq. (\ref{eq:cross_section_tilde}) with respect to $\nu$, one gets
\begin{align*}
c \nu \int d \bq \tilde{\sigma}(\bq ; \bp) q_n & = - \nu \sum_{m=1}^{d} \partial_{p_m} D^\nu_{nm}(\bp) + \mathcal{O}(\nu^2) \\
c \nu^2 \int d \bq \tilde{\sigma}(\bq ; \bp) q_n q_m & = 2 \nu D^\nu_{nm}(\bp) + \mathcal{O}(\nu^2)
\end{align*}
with $\boldsymbol{D}^\nu (\bp) = - (c/2) \int_{\mathbb{R}} \nabla \otimes \nabla R^\nu(s \bp) ds$. Here, $\int_{\mathbb{R}} \nabla R^\nu (s \bp) ds$ vanishes because of the point symmetry in $R^\nu$. Inserting these expressions into \eqref{eq:diffusion_1}, one obtains
\begin{align} \label{eq:diffusion_final}
& \int d \bp_1 d \bp_2 f(\bp_1)\Sigma(\bp_1,\bp_2)g(\bp_2) \nonumber \\
= &  - c \nu \int d \bp fn \sum_{n=1}^d \sum_{m=1}^d \left\{ n\partial_{p_{n}} D^\nu_{nm} \partial_{p_{m}}g + 2D^\nu_{nm} \partial_{p_{n}} n \partial_{p_{m}} g + n D^\nu_{nm} \partial_{p_{n}} \partial_{p_{m}} g \right\}  \nonumber \\
+ & \mathcal{O}(\nu^2) .
\end{align}
Evidently from this result, the dominant term is of order $\mathcal{O}(\nu)$ in the present scaling. Therefore, in order for diffusion in wave-vector space to be comparable to the free propagation in position space, one needs to rescale the position coordinate. However, for the sake of simplicity here, we shall set $\nu = 1$ while ignoring $\mathcal{O}(\nu^2)$ terms. Consequently, after reorganizing \yo{derivatives of products} and performing integration by parts, we confirm that \eqref{eq:diffusion_final} transforms into Eq.~\eqref{eq:kernel_Sigma_diffusive_limit}.

\bibliography{Linear_Schrodinger}

\end{document}